\documentclass[a4paper]{article}
\usepackage{indentfirst}
\usepackage{booktabs}
\usepackage{hyperref}
\usepackage{multirow}
\usepackage{latexsym,bm,amsmath,amssymb}
\usepackage{graphicx}
\usepackage{epsfig}
\usepackage{subfigure}
\usepackage{cite}
\usepackage{color}
\usepackage[top=1in, bottom=1in, left=1.25in, right=1.25in]{geometry}
\usepackage{ulem}
\usepackage{epstopdf}

\def\bea{\arraycolsep .1em \begin{eqnarray}}
\def\eea{\end{eqnarray}}

\def\s0#1#2{\mbox{\small{$ \frac{#1}{#2} $}}}
\def\0#1#2{\frac{#1}{#2}}

\begin{document}

\setcounter{topnumber}{10}
\setcounter{totalnumber}{50}
\title{ On leptonic width of $X(4260)$ }
\maketitle
%%%%%%%%%%%%%%%%%%%%%%%%%%%%

%%%%%%%%%%%%%%%%%%%%%%%%%%%%
\begin{center}
{\sc Qin-Fang Cao$^{\dagger}$,\,\,\,Hong-Rong Qi$^\ddagger$,\,\,\, Guang-Yi Tang,$^*$\,\,\,
Yun-Feng Xue,$^\dagger$\,\,\, Han-Qing Zheng$^{\dagger\,,\star\,,}$
%\footnote{e-mail address:zhenghq@pku.edu.cn}
}
\\
\vspace{0.5cm}

\noindent{\small{$^\dagger$ \it  Department of Physics and State Key
Laboratory of Nuclear Physics and Technology,
 Peking University, Beijing 100871, People's Republic of China}}\\
\noindent{\small{$^\ddagger$ \it Department of engineering physics, Tsinghua University, Beijing 100084, People's Republic of China}}\\
\noindent{\small{$^*$ \it  Institute of High Energy Physics, Beijing 100049, People's Republic of China}}\\
\noindent{\small{$^\star$ \it Collaborative Innovation Center of Quantum Matter, Beijing 100871, People's Republic of China}}
\end{center}
\begin{abstract}
 New measurements on cross sections in $e^+e^-\to J/\psi\pi^+\pi^-$, $h_c\pi^+\pi^-$, $D^0D^{*-}\pi^++c.c.$, $\psi(2S)\pi^+\pi^-$, $\omega\chi_{c0}$ and $J/\psi\eta$ channels have been carried out by BESIII, Belle and BABAR experiments, as well as in the $D_s^{*+}D_s^{*-}$ channel. We perform  extensive numerical analyses by combining all these data available, together with those in $D^+D^{*-}+c.c.$ and $D^{*+}D^{*-}$ channels. Though the latter show no evident peak around $\sqrt{s}=4.230$ GeV, the missing $X(4260)$ is explained as that it is concealed by the interference effects of the well established charmonia $\psi(4040)$, $\psi(4160)$ and $\psi(4415)$. Our analyses reveal that the leptonic decay width of $X(4260)$ ranges from $\mathcal{O}(10^2)$ eV to $\mathcal{O}(1)$ keV, and hence it is probably explained in the conventional quark model picture. That is, the $X(4260)$ may well be interpreted as a mixture of $4^3S_1$ and $3^3D_1$ ($2^3D_1$) states.
\end{abstract}

\section{Introduction}\label{intro}

 The $X(4260)$  resonance
 established by BABAR Collaboration in initial-state radiation (ISR) process
 $e^+e^-\rightarrow \gamma_{ISR}J/\psi\pi^+\pi^-$ in
 year 2005\cite{Babary4260} (subsequently confirmed by CLEO\cite{CLEOy4260a} and
 Belle\cite{Belley4260}), has attracted much attention since then.
 The mass and width of this resonance are given with
 $M=4230\pm8$ MeV and $\Gamma=55\pm19$ MeV~\cite{PhysRevD.98.030001}, respectively, and $\Gamma_{ee}\times\mathrm{Br}(J/\psi\pi\pi)=9.7\pm1.1$ eV~\cite{Belley4260} or $9.2\pm1.5$ eV~\cite{BaBar2012}.

\begin{figure}
\centering
\vspace{0cm}
\includegraphics[width=0.8\textwidth]{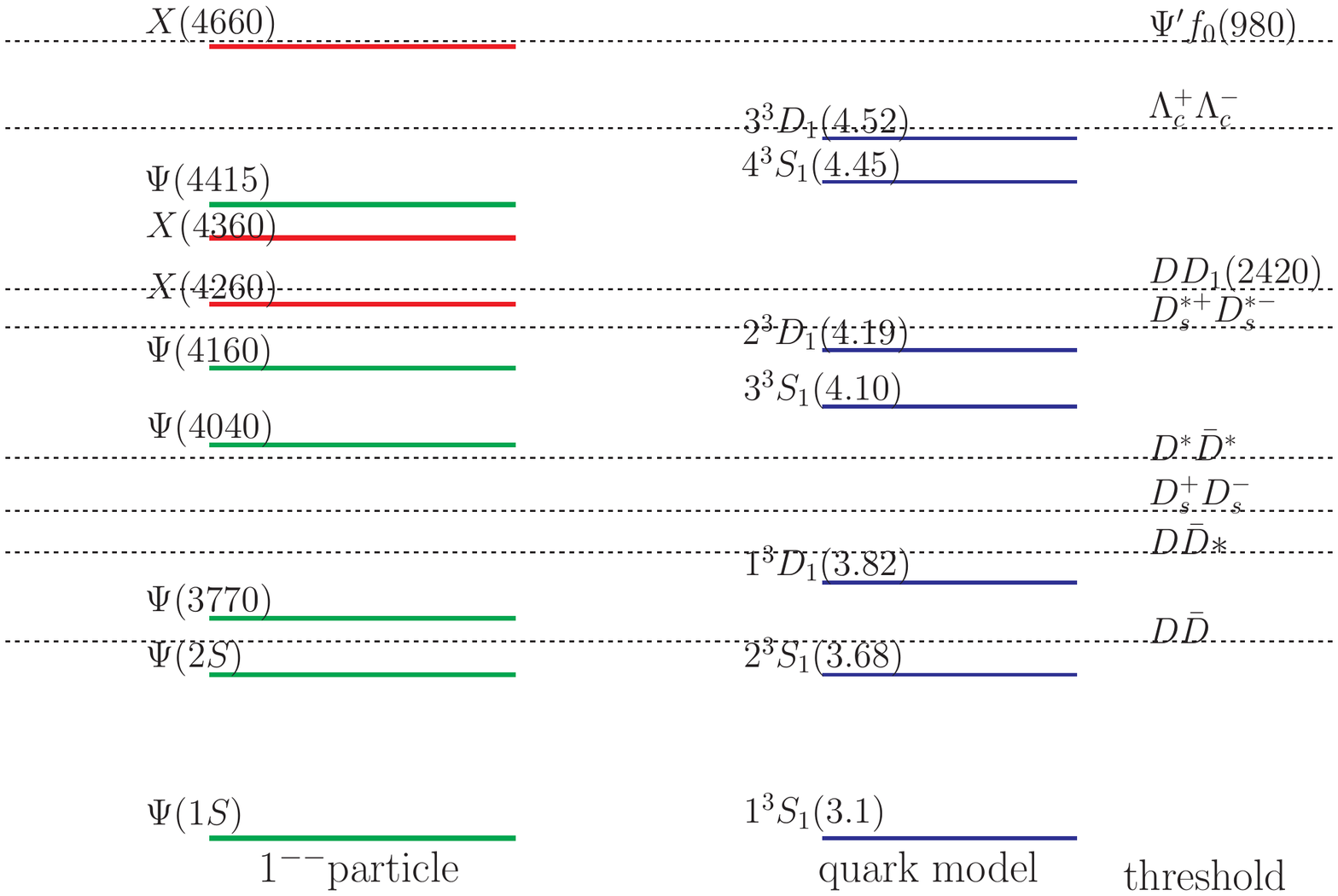}
\vspace{-2cm}
  \caption{\small $X(4260)$ and nearby resonances from naive quark model calculation\cite{nqm}. }\label{threshold}
 \end{figure}

\begin{figure}
 \centering
 \vspace{1cm}
 \includegraphics[width=0.8\textwidth]{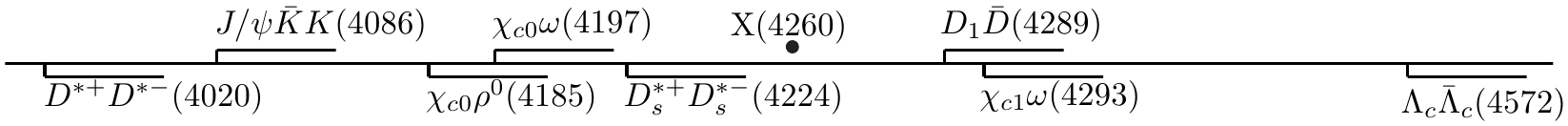}
  \caption{\small Location of $X(4260)$ and nearby thresholds. }\label{X4260threh}
 \end{figure}

The property of $X(4260)$ becomes a very interesting topic since its discovery, because it is generally thought that there are not enough unassigned
vector states in charmonium spectrum (taking into account the recently reported $X(4360)$, $X(4630)$/$X(4660)$ states), according to the naive quark model predictions~\cite{nqm}. The
only such $1^{--}$ states expected up to $4.4$ GeV are generally $1S$, $2S$, $1D$,
$3S$, $2D$ and $4S$, and they seem to be well established~\cite{review}
-- the situation is depicted in Figure~\ref{threshold}.
%%%%%%%%%%%%%%%%%%%%%%%%%%
%%%%%%%%%%%%%%%%%%%%%%%%%%%%%%%%%%%%%%%%%%%%%%%%%%%%%%%%%%%
It is noticed that above $D\bar{D}$ threshold the number of
$1^{--}$ states given by quark model prediction is inconsistent
with that given by experiments. It is considered  that the
discrepancy between the naive quark model prediction and the
observed spectrum is ascribed, at least partially, to the existence
of many open charm thresholds, since the latter will distort the
spectrum (see related discussions in, for example, Refs.~\cite{Zhou:2011sp,Zhou:2013ada}). The situation is depicted in Figure~\ref{X4260threh}.

Because of the situation as described above many studies
have been devoted to the investigation of $X(4260)$.\footnote{Here we apologize for only able to provide an incomplete list of references.} The suggestions given by these studies may be classified into three types: hadronic molecule~\cite{Xue:2017xpu,Qin:2016spb,Ding:2008gr,Close:2009ag,Li:2013bca,YDChen2013}; $c\bar c $ state~\cite{Wang:2019mhs,Huang:2019agb,Chen:2017uof,Li:2009zu,LlanesEstrada:2005hz};  hybrid state~\cite{Kou:2005gt,Zhu:2005hp}; or non-resonant enhancement~\cite{Coito:2019cts}.

The open charm channels such as $D \bar D,D\bar
D^*,D^*\bar D^*$ do not seem to be found in the final states of $X(4260)$
decays~\cite{Babar2009,PhysRevD.77.011103,PhysRevLett.98.092001}. If this is indeed the case, then it would make $X(4260)$ even more mysterious,
since the $J/\psi\pi\pi$ channel would become a very important, if not the dominant one. Hence the leptonic width $\Gamma_{e^+e^-}$ of $X(4260)$ would become very small, making it even harder to be understood as a conventional $1^{--}$ $\bar cc$ state, since the nearby $4 ^3S_1$ state is expected to have a leptonic width at the order $\sim 10^3$ eV~\cite{Li:2009zu}.

In a previous publication, we have also studied the $X(4260)$ issue and suggested that there could be a sizable $\omega\chi_{c0}$ coupling~\cite{Tang2015}, later confirmed by
experimental researches~\cite{Ablikim:2014qwy}.
At the same time, a very small $\Gamma_{e^+e^-}$ was found $\sim25$eV. However,  many new experimental results have appeared since then, measured by BESIII, Belle and BABAR experiments, such as $e^+e^-\to J/\psi\pi^+\pi^-$~\cite{Ablikim:2016qzw,BaBar2012,Belley4260}, $h_c\pi^+\pi^-$~\cite{BESIII:2016adj}, $D^0D^{*-}\pi^++c.c.$~\cite{Ablikim:2018vxx}, $\psi(2S)\pi^+\pi^-$~\cite{Ablikim:2017oaf,Wang:2014hta,Lees:2012pv}, $\omega\chi_{c0}$~\cite{Ablikim:2019apl,Ablikim:2015uix}, $J/\psi\eta$~\cite{Ablikim:2015xhk,Wang:2012bgc} and $D_s^{*+}D_s^{*-}$~\cite{Ds*Ds*}. Hence the analyses of Ref.~\cite{Tang2015} urgently need to be upgraded. Among all it is worthwhile mentioning the $D_s^{*+}D_s^{*-}$ data near the $X(4260)$ region~\cite{Ds*Ds*}, which indicates a strong enhancement of events above the $D_s^{*+}D_s^{*-}$ threshold. If this is true, our analyses show that it decisively changes our previous understandings on $X(4260)$ resonance: It could probably be described by the conventional $4 ^3S_1$ state heavily renormalized by the $D_s^{*+}D_s^{*-}$ continuum (maybe a small mixing with the $3^3D_1$ state as well). {If the $D_s^{*+}D_s^{*-}$ data are excluded from the fit, however, the final result on $\Gamma_{e^+e^-}$ can still be at least $\mathcal{O}(10^2)$ eV, i.e., much larger comparing with that of Ref.~\cite{Tang2015}, owing to other new data available as mentioned above. As a consequence, the $X(4260)$ resonance may still be considered as a mixture of $3^3D_1$ and $4 ^3S_1$ states, i.e., a conventional $\bar cc$ resonance.}

 The paper is organized as follows: This section~\ref{intro} is the introduction. A detailed description of hadronic cross sections of $e^+e^-$ annihilation will be given in section~\ref{theory}. In section~\ref{numerical}, combined fits to the hidden charm and open charm decay channels are
performed, with two scenarios: one  includes the $D_s^{*+}D_s^{*-}$ cross section data and the other  does not. We leave physical discussions and conclusions in section~\ref{conclusion}.

\section{Theoretical Discussions}\label{theory}
To begin with, the $X(4260)$ propagator
is written in the following form:

\begin{equation}
\begin{split}
\frac{1}{D_X(s)}=\frac{1}{s-M_X^2+i\sqrt{s}\Gamma_{tot}(s)},
\end{split}
\end{equation}
where $M_X$ represents the mass of $X(4260)$ and $\Gamma_{tot}(s)$ is the total momentum dependent width comprising of all partial ones:~\footnote{We do not consider a term $\propto \Gamma_{DD_1}$ here, as it is found vanishing in Ref.~\cite{Tang2015}. Also the experimental data in $Z_c\pi$ channel is absent. Since the $Z_c(3900)$ state is identified as a $DD^*$ molecule~\cite{ZcGong, ZcPang}, the experimental branching ratio  is naturally expected to be small.}
\begin{equation}\label{eq:gammatotal}
\begin{split}
\Gamma_{tot}(s)=&\Gamma_{J/\psi\pi\pi}(s)+\Gamma_{h_c\pi\pi}(s)+\Gamma_{D\bar D^{*}\pi}(s)+
\Gamma_{\psi(2S)\pi\pi}(s)+\Gamma_{\omega\chi_{c0}}(s)+\Gamma_{J/\psi\eta}(s)\\
&+\Gamma_{D_s^{*+}D_s^{*-}}(s)+\Gamma_{D\bar D}(s)+\Gamma_{D\bar D^*}(s)+\Gamma_{D^*\bar D^*}(s)+\Gamma_0.\\
\end{split}
\end{equation}
Considering the isospin symmetry, it is noticed that $\Gamma_{J/\psi\pi\pi}(s)=\frac{3}{2}\Gamma_{J/\psi\pi^+\pi^-}(s)$,
$\Gamma_{h_c\pi\pi}(s)=\frac{3}{2}\Gamma_{h_c\pi^+\pi^-}(s)$, $\Gamma_{D\bar D^{*}\pi}(s)=3\Gamma_{D^0D^{*-}\pi^++c.c.}(s)$, $\Gamma_{\psi(2S)\pi\pi}(s)=
 \frac{3}{2}\Gamma_{\psi(2S)\pi^+\pi^-}(s)$, $\Gamma_{D\bar D}=2\Gamma_{D^+D^-}$, $\Gamma_{D\bar D^*}=2\Gamma_{D^+D^{*-}+c.c.}$ and $\Gamma_{D^*\bar D^*}=2\Gamma_{D^{*+}D^{*-}}$. \footnote{Hereafter, the notation $D^0D^{*-}\pi^+$ and $D^+D^{*-}$ indicate $D^0D^{*-}\pi^++c.c.$ and $D^+D^{*-}+c.c.$ for simplicity, respectively. }

As for the $J/\psi\pi^+\pi^-$, $h_c\pi^+\pi^-$, $D^0D^{*-}\pi^+$ and $\psi(2S)\pi^+\pi^-$ channels, the three body partial decay width takes the standard form~\cite{PhysRevD.98.030001},\footnote{Here we use the subscript ``$f$'' to represent various three body decay final states.}
\begin{equation}\label{eq:gamma3}
d\Gamma_{f}=\frac{1}{(2\pi)^5}\frac{1}{16M_{X}^2}|\mathcal{M}|^2|p_1^*||p_3|dm_{12}d\Omega_1^*d\Omega_3,
\end{equation}
where $p_{ij}=p_i+p_j$, $m_{ij}^2=p_{ij}^2$, $|p_1^*|$, $\Omega_1^*$ are the momentum and angle of particle 1 in the rest frame of the system of particle 1 and 2, respectively; $\Omega_3$ is the angle of particle 3 in the rest frame of particle $M_X$; and $|p_1^*|$, $|p_3|$ are defined as
\begin{equation}
|p_1^*|=\frac{[(m_{12}^2-(m_1+m_2)^2)(m_{12}^2-(m_1-m_2)^2)]^{1/2}}{2m_{12}},
\end{equation}
\begin{equation}
|p_3|=\frac{[(M_X^2-(m_{12}+m_3)^2)(M_X^2-(m_{12}-m_3)^2)]^{1/2}}{2M_X}.
\end{equation}
With respect to $J/\psi\pi^+\pi^-$ channel, the amplitude in Eq.~(\ref{eq:gamma3}) is calculated using the effective Lagrangian as Eq.~(6) in {Ref.~\cite{Tang2015}} to describe the $X(4260)\to J/\psi\pi^+\pi^-$ process and the interaction coefficients $h_i(i=1,2,3)$ are listed in Tables~\ref{tab:coupling1} and~\ref{tab:coupling2}. Since the production ratio $R=(\sigma(e^+e^-\to Z_c(4020)^\mp\pi^\pm\to h_c\pi^+\pi^-)/\sigma(e^+e^-\to h_c\pi^+\pi^-))<20\%$~\cite{Ablikim:2013wzq}, $g_{h_c\pi^+\pi^-}$ can be used to describe the invariant amplitude in Eq.~(\ref{eq:gamma3}) when ignoring the resonance structure in the $X(4260)$ decay. Analogously, $g_{D^0D^{*-}\pi^+}$, $g_{\psi(2S)\pi^+\pi^-}$ can also be introduced as the invariant amplitude in Eq.~(\ref{eq:gamma3}) for $D^0D^{*-}\pi^+$ and $\psi(2S)\pi^+\pi^-$ channels for simplicity. Besides, in the analyses below, it is noticed that the decay ratios of these channels account for only a tiny share. Hence, it is feasible to employ $g_{h_c\pi^+\pi^-}$, $g_{D^0D^{*-}\pi^+}$ and $g_{\psi(2S)\pi^+\pi^-}$ for simplicity.

In addition, the two body decay widths take the following simple forms:
\begin{equation}\label{eq:gamma2}
\begin{split}
&\Gamma_{\omega\chi_{c0}}(s)=g_{\omega\chi_{c0}}k_{\omega\chi_{c0}},\quad\Gamma_{J/\psi\eta}(s)=g_{J/\psi\eta}k^3_{J/\psi\eta}, \quad\Gamma_{D_s^{*+}D_s^{*-}}(s)=g_{D_s^{*+}D_s^{*-}}k^3_{D_s^{*+}D_s^{*-}},\\
&\Gamma_{D^+D^-}(s)=g_{D^+D^-}k_{D^+D^-}^3,\quad\Gamma_{D^+D^{*-}}(s)=g_{D^+D^{*-}}k_{D^+D^{*-}}^3,\\
&\Gamma_{D^{*+}D^{*-}}(s)=g_{D^{*+}D^{*-}}k^3_{D^{*+}D^{*-}}.
\end{split}
\end{equation}
In above $k_{\omega\chi_{c0}}$, $k_{J/\psi\eta}$, $k_{D_s^{*+}D_s^{*-}}$, $k_{D^+D^-}$, $k_{D^+D^{*-}}$ and $k_{D^{*+}D^{*-}}$ are three momentums of $\omega\chi_{c0}$, $J/\psi\eta$, $D_s^{*+}D_s^{*-}$, $D^+D^-$, $D^+D^{*-}$ and $D^{*+}D^{*-}$ in $X(4260)$ rest frame, respectively. Further, except for the channels just discussed, there could be other channels with lower thresholds, for these channels we use a constant width $\Gamma_0$ to describe the overall effects.

Because the quantum number of $X(4260)$ is $J^{PC}=1^{--}$, the interaction between $X(4260)$ and photon can be written as
\begin{equation}
\mathcal{L}_{\gamma X}=\frac{g_0}{M_X}X_{\mu\nu}F^{\mu\nu},
\end{equation}
where $X_{\mu\nu}=\partial_\mu V_\nu-\partial_\nu V_\mu$ and $V^\mu$ represents the $X(4260)$ field, $F^{\mu\nu}=\partial^\mu A^\nu-\partial^\nu A^\mu$ describes the photon field. $g_0$ is the coefficient between the photon and $X(4260)$. So the decay width of $X(4260)\rightarrow e^+e^-$ reads
\begin{equation}\label{eq:gammaee}
\Gamma_{e^+e^-}=\frac{4\alpha}{3}\frac{g_0^2}{M_X}.
\end{equation}

In $h_c\pi^+\pi^-$, $D^0D^{*-}\pi^+$, $\psi(2S)\pi^+\pi^-$, $\omega\chi_{c0}$, $J/\psi\eta$, $D_s^{*+}D_s^{*-}$, $D^+D^{*-}$ and $D^{*+}D^{*-}$ channels, using narrow width approximation, the cross section formulae take the form
\begin{equation}\label{eq:sigmaf}
\sigma_{e^+e^-\rightarrow X(4260)\rightarrow f}=\frac{3\pi}{k^2}|\frac{\sqrt{s\Gamma_{ee}\Gamma_f}}{s-M_X^2+i\sqrt{s}\Gamma_{tot}(s)}+{\sum_i \frac{c_ie^{i\phi_i}}{s-M_i^2+i\sqrt{s}\Gamma_i}+\tilde{c}} \,|^2,
\end{equation}
where $k$ is the 3-momentum of incoming electron in $c.m.$ frame, $\Gamma_{ee}$ follows Eq.~(\ref{eq:gammaee}) and $\Gamma_f$ takes the form of Eq.~(\ref{eq:gamma3}) and Eq.~(\ref{eq:gamma2}). The term parameterized as a resonance propagator with a mass $M_i$ and width $\Gamma_i$ and the complex constant $\tilde c$ play the role of a  background in each decay channel here, as will be declared in details in the forthcoming section~\ref{numerical}.

\section{Numerical Analyses and Discussions}\label{numerical}
In section~\ref{fitwithdsds}, we will perform comprehensive fits to relevant data available in the vicinity of $X(4260)$, which include the $J/\psi\pi^+\pi^-$~\cite{Ablikim:2016qzw,BaBar2012,Belley4260}, $h_c\pi^+\pi^-$~\cite{BESIII:2016adj}, $D^0D^{*-}\pi^+$~\cite{Ablikim:2018vxx}, $\psi(2S)\pi^+\pi^-$~\cite{Ablikim:2017oaf,Wang:2014hta,Lees:2012pv}, $\omega\chi_{c0}$~\cite{Ablikim:2019apl,Ablikim:2015uix},  $J/\psi\eta$~\cite{Ablikim:2015xhk,Wang:2012bgc}, $D_s^{*+}D_s^{*-}$~\cite{Ds*Ds*}, together with the previous $D^+D^{*-}$ and $D^{*+}D^{*-}$ data in Ref.~\cite{PhysRevLett.98.092001}.
To be cautious, for the reason as already mentioned previously, the fit without $D_s^{*+}D_s^{*-}$ cross section data is also performed in section~\ref{fitwithoutdsds}. Different results are carefully compared and discussed, and we believe that a clearer understanding on the nature of $X(4260)$ emerges.

\subsection{The fit with $D_s^{*+}D_s^{*-}$ cross section data}\label{fitwithdsds}
\subsubsection{The $J/\psi\pi^+\pi^-$ process}
\begin{figure}[htbp]
\center
\subfigure[$J/\psi \pi^+\pi^-$]{
\scalebox{1.2}[1.2]{\includegraphics[width=0.25\textwidth]{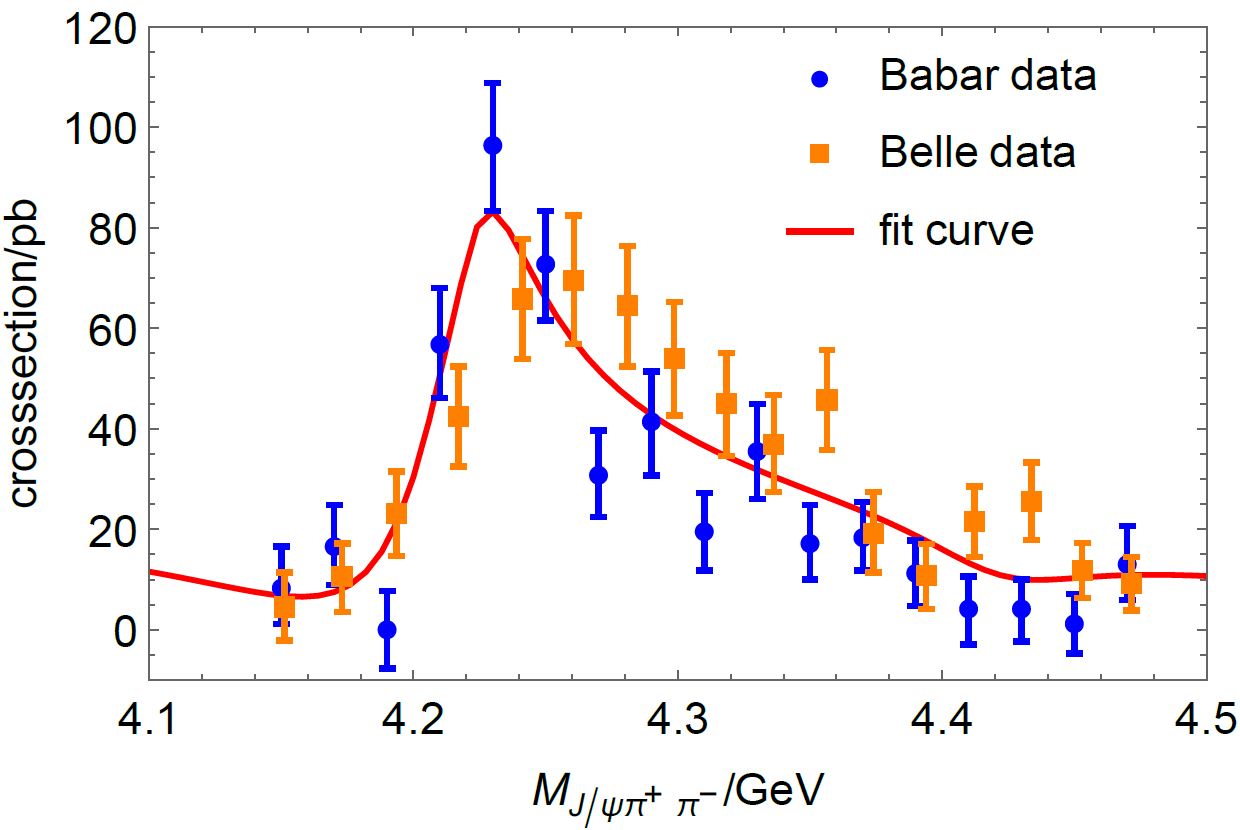}}
\label{fig:jpsipipi1}
}
\subfigure[$J/\psi \pi^+\pi^-$]{
\scalebox{1.2}[1.2]{\includegraphics[width=0.25\textwidth]{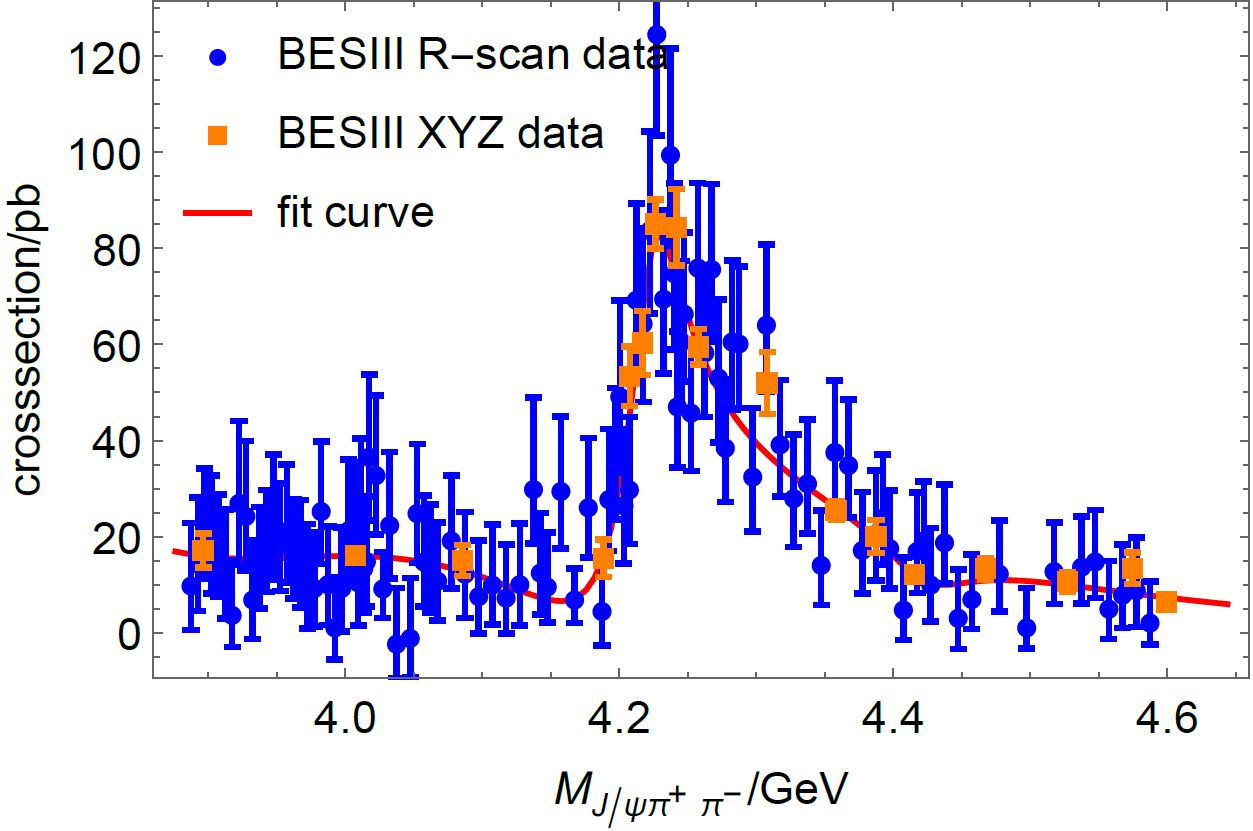}}
\label{fig:jpsipipibes1}
}
\subfigure[$D^0D^{*-}\pi^+$]{
\scalebox{1.2}[1.2]{\includegraphics[width=0.25\textwidth]{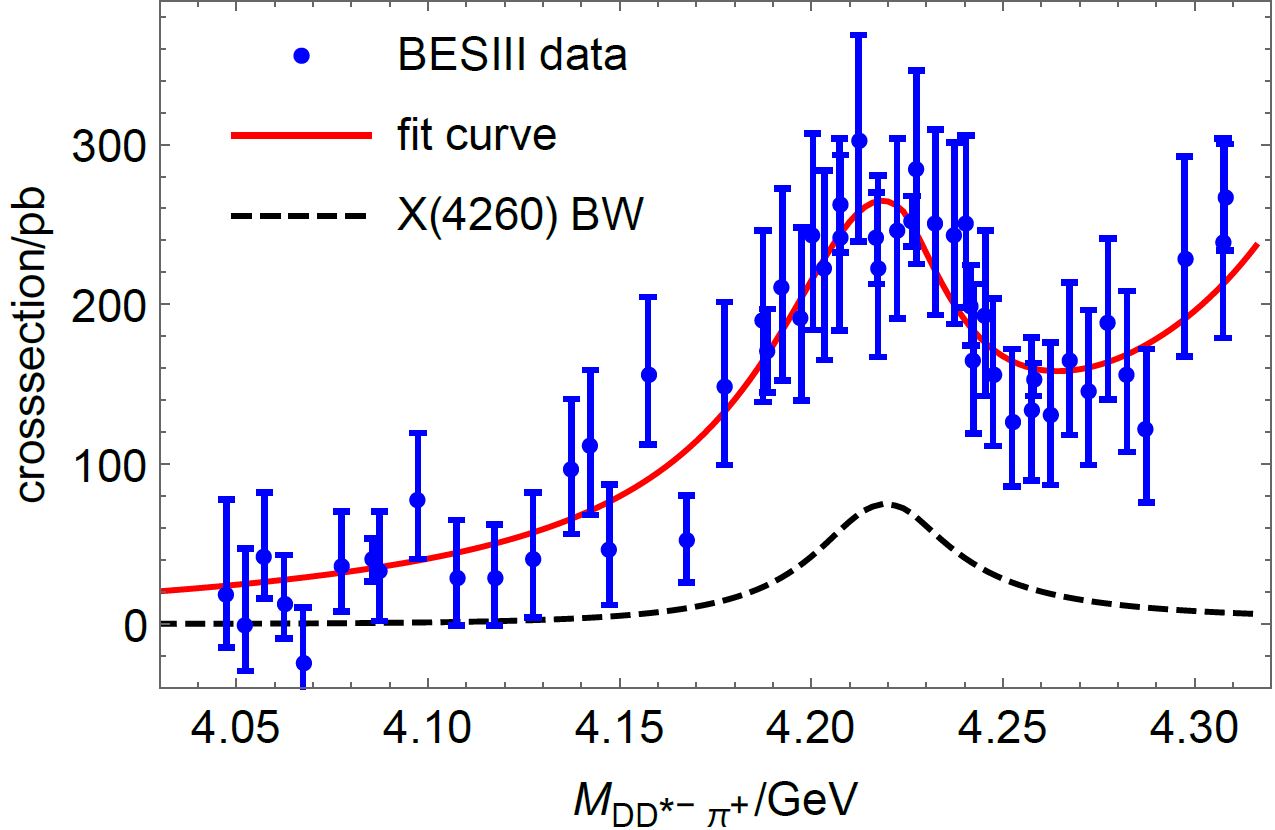}}
\label{fig:ddpi1}
}
\subfigure[$h_c\pi^+\pi^-$]{
\scalebox{1.2}[1.2]{\includegraphics[width=0.25\textwidth]{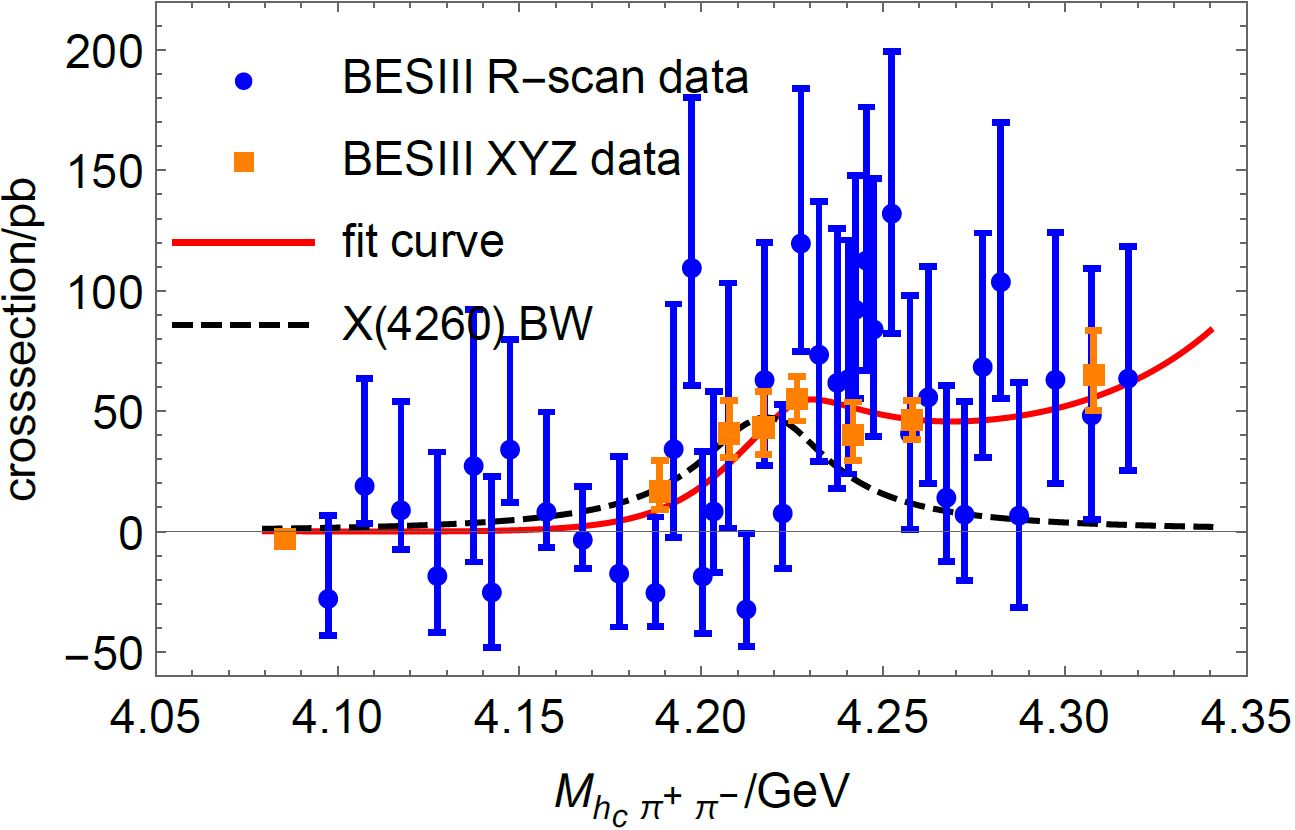}}
\label{fig:hcpipi1}
}
\subfigure[$\psi(2S)\pi^+\pi^-$]{
\scalebox{1.2}[1.2]{\includegraphics[width=0.25\textwidth]{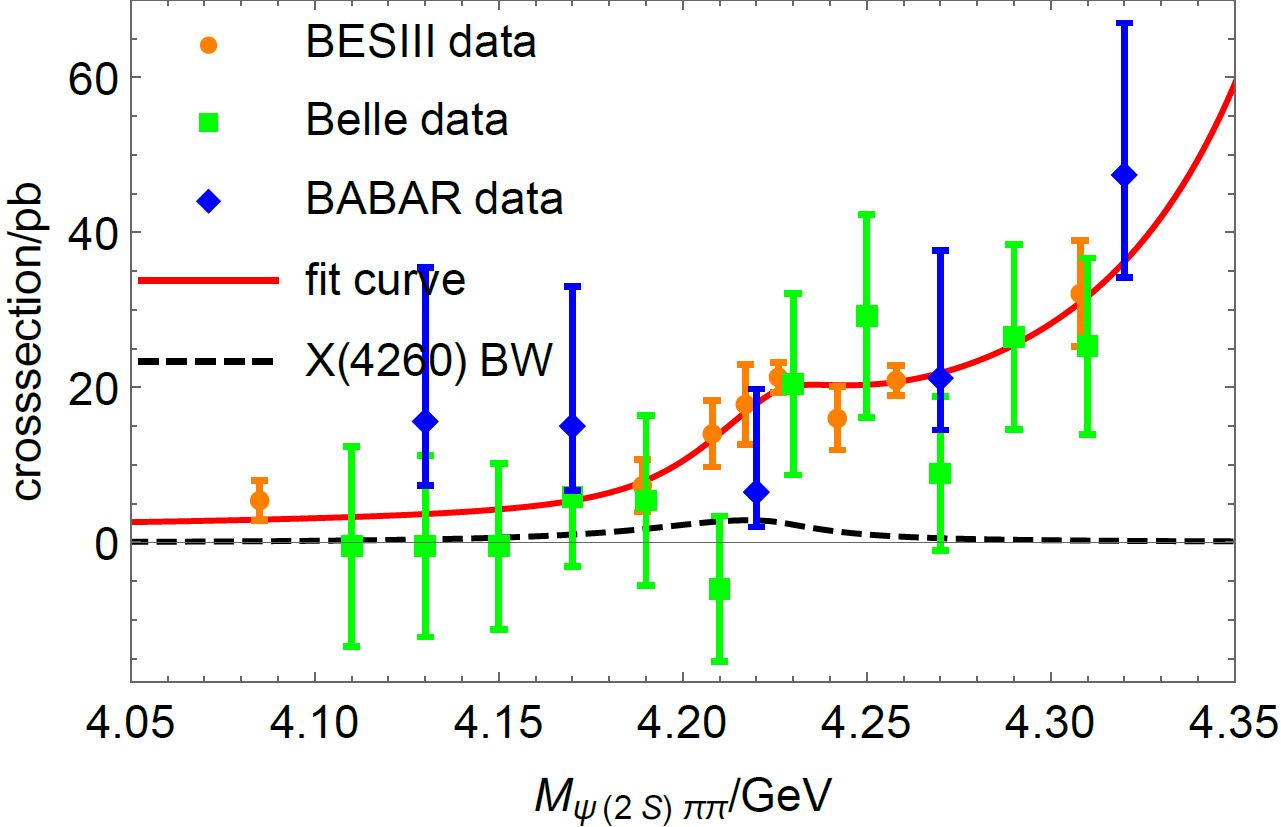}}
\label{fig:psi2spipi1}
}
\subfigure[$\omega\chi_{c0}$]{
\scalebox{1.2}[1.2]{\includegraphics[width=0.25\textwidth]{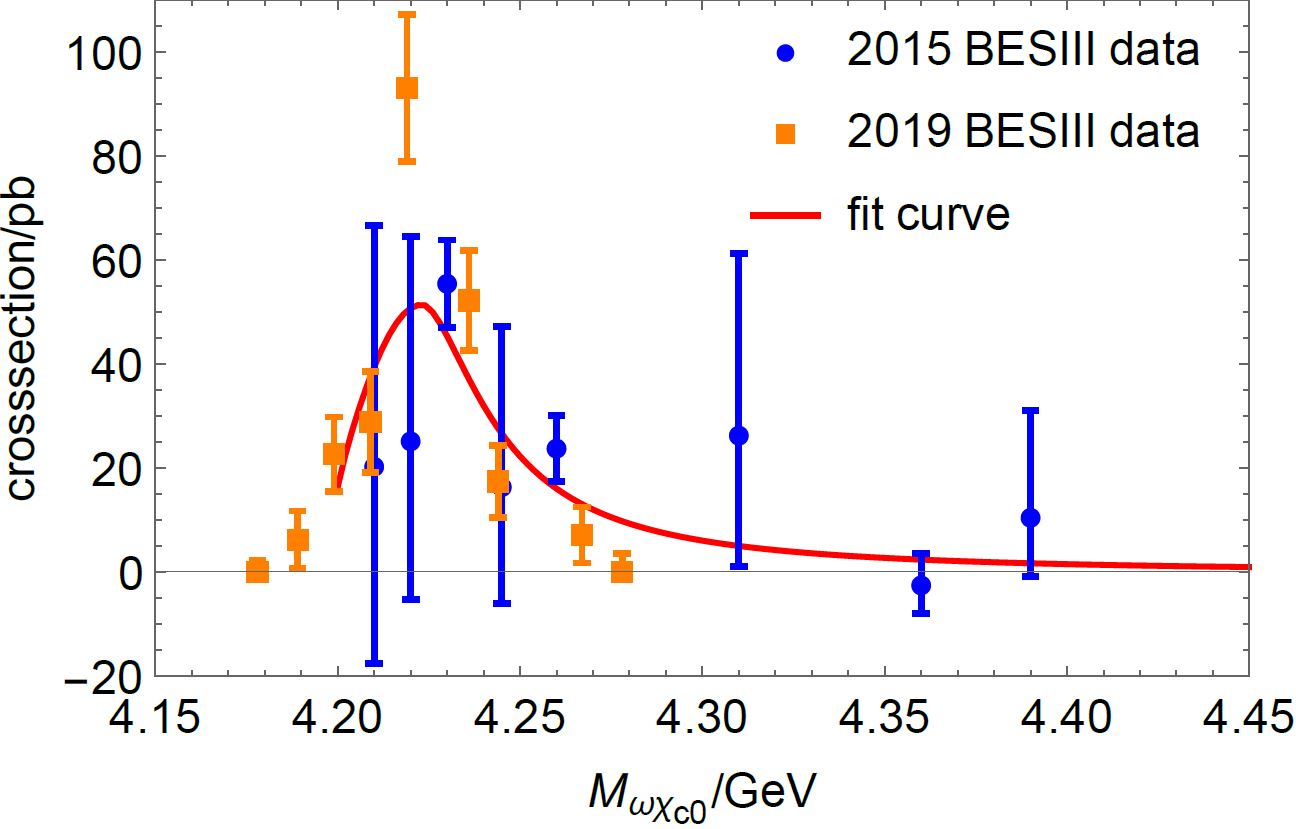}}
\label{fig:omegakaic01}
}
\subfigure[$J/\psi\,\eta$ ]{
\scalebox{1.2}[1.2]{\includegraphics[width=0.25\textwidth]{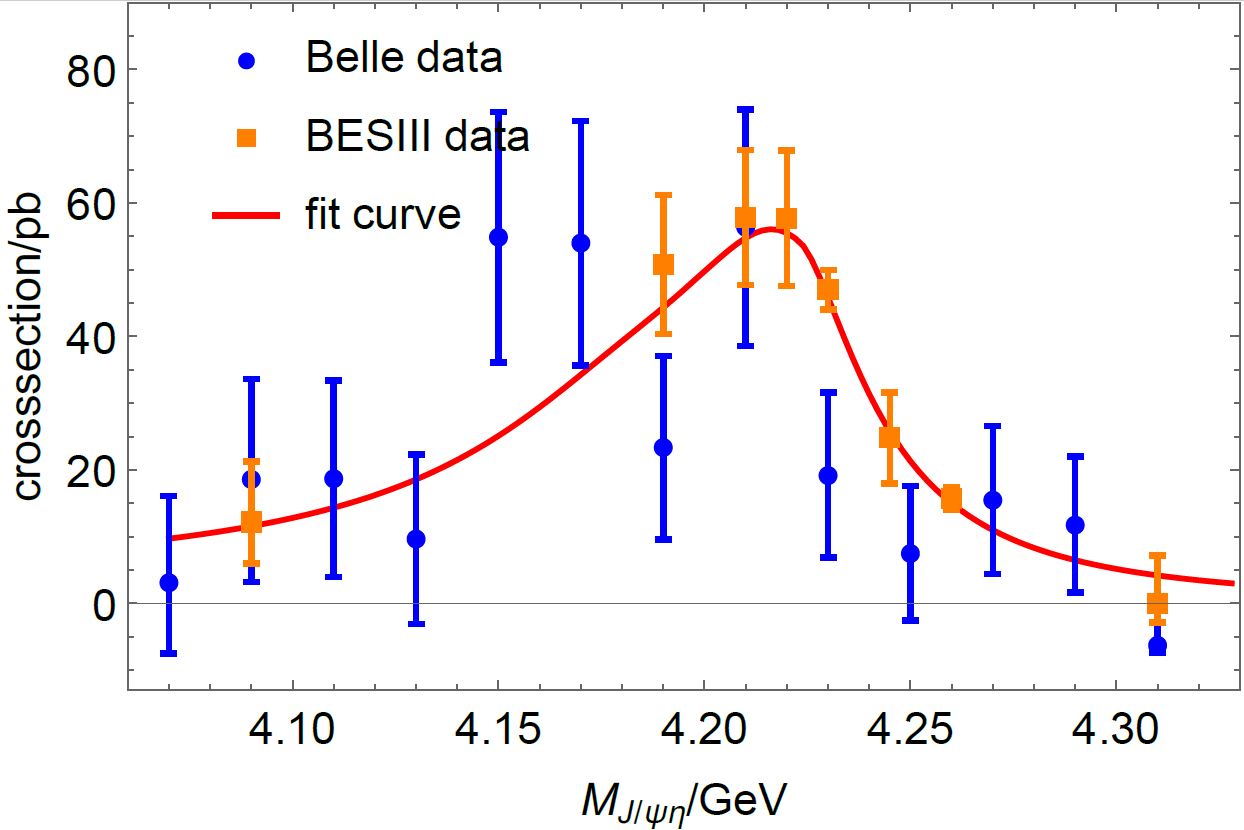}}
\label{fig:jpsieta1}
}
\subfigure[$D_s^{*+}D_s^{*-}$ ]{
\scalebox{1.2}[1.2]{\includegraphics[width=0.25\textwidth]{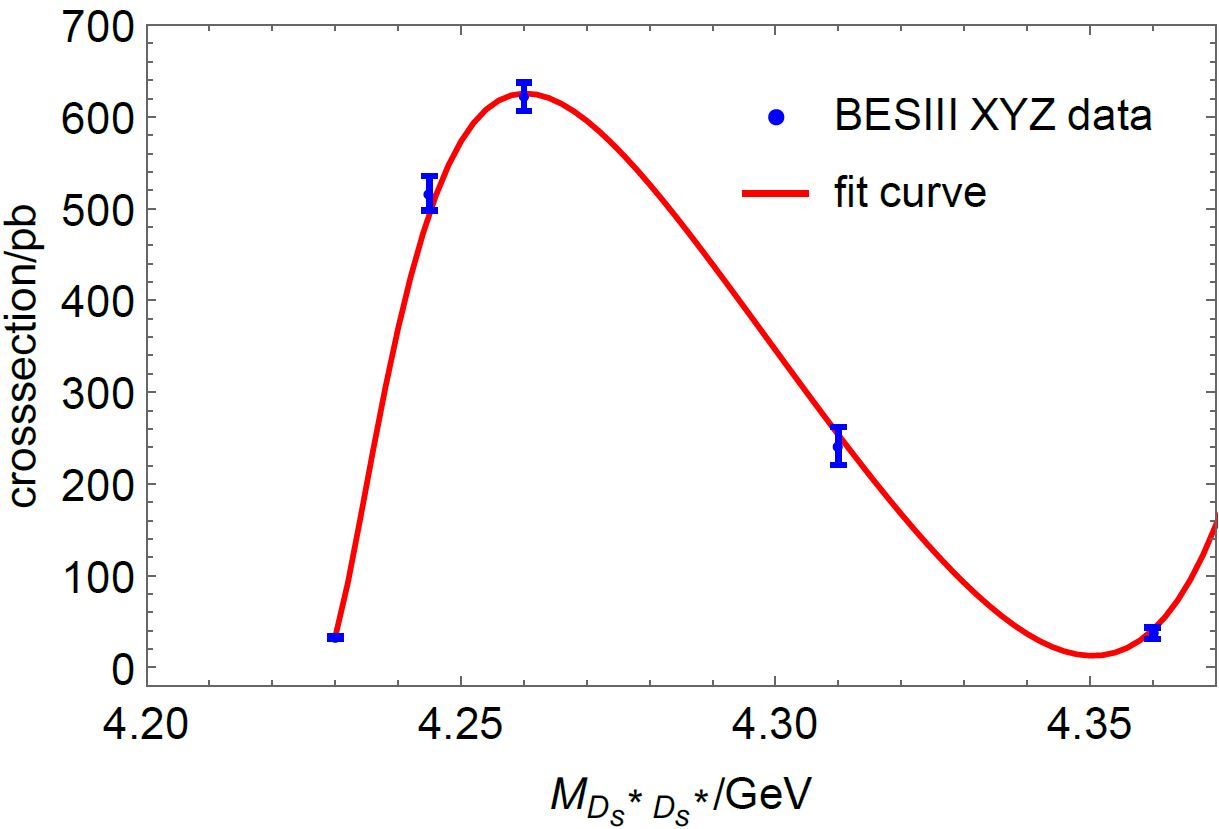}}
\label{fig:dsds1}
}
\subfigure[$D^+D^{*-}$]{
\scalebox{1.2}[1.2]{\includegraphics[width=0.25\textwidth]{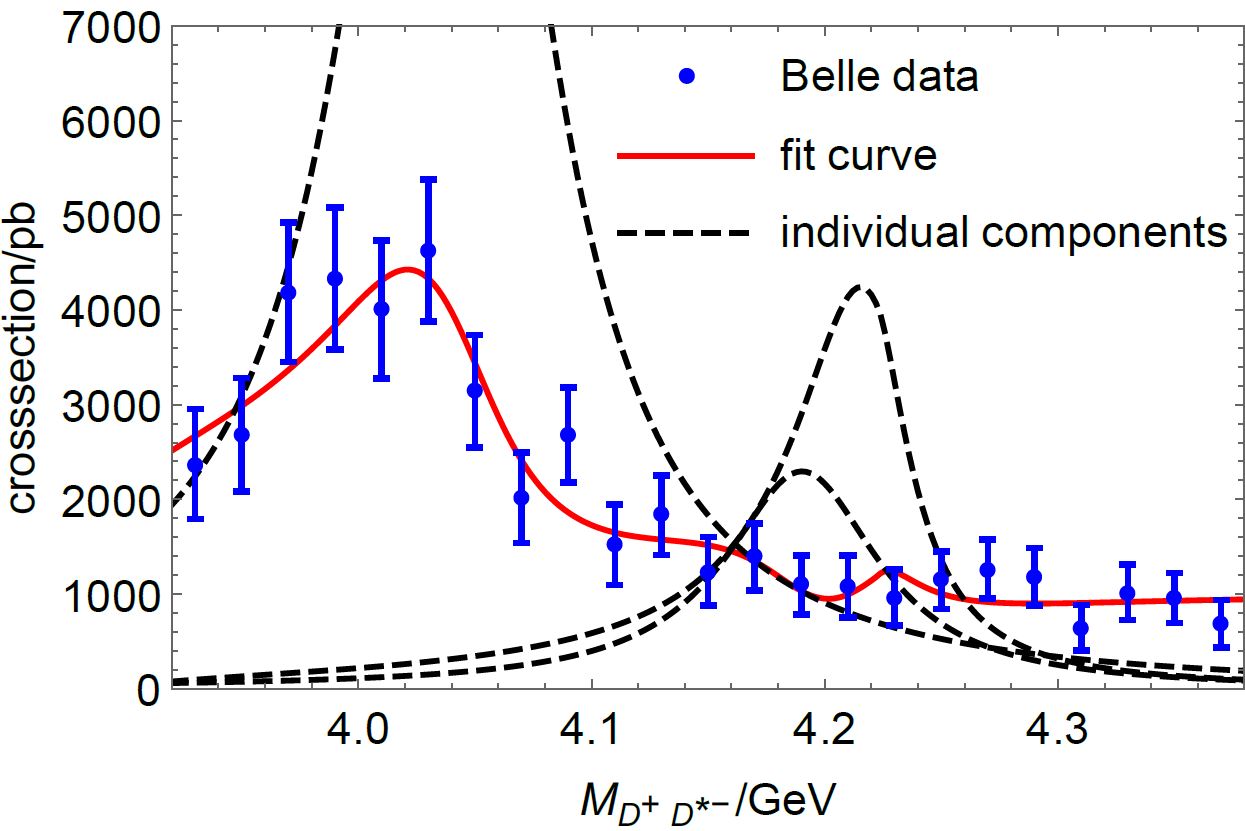}}
\label{fig:ddstar1}
}
\subfigure[$D^{*+}D^{*-}$]{
\scalebox{1.2}[1.2]{\includegraphics[width=0.25\textwidth]{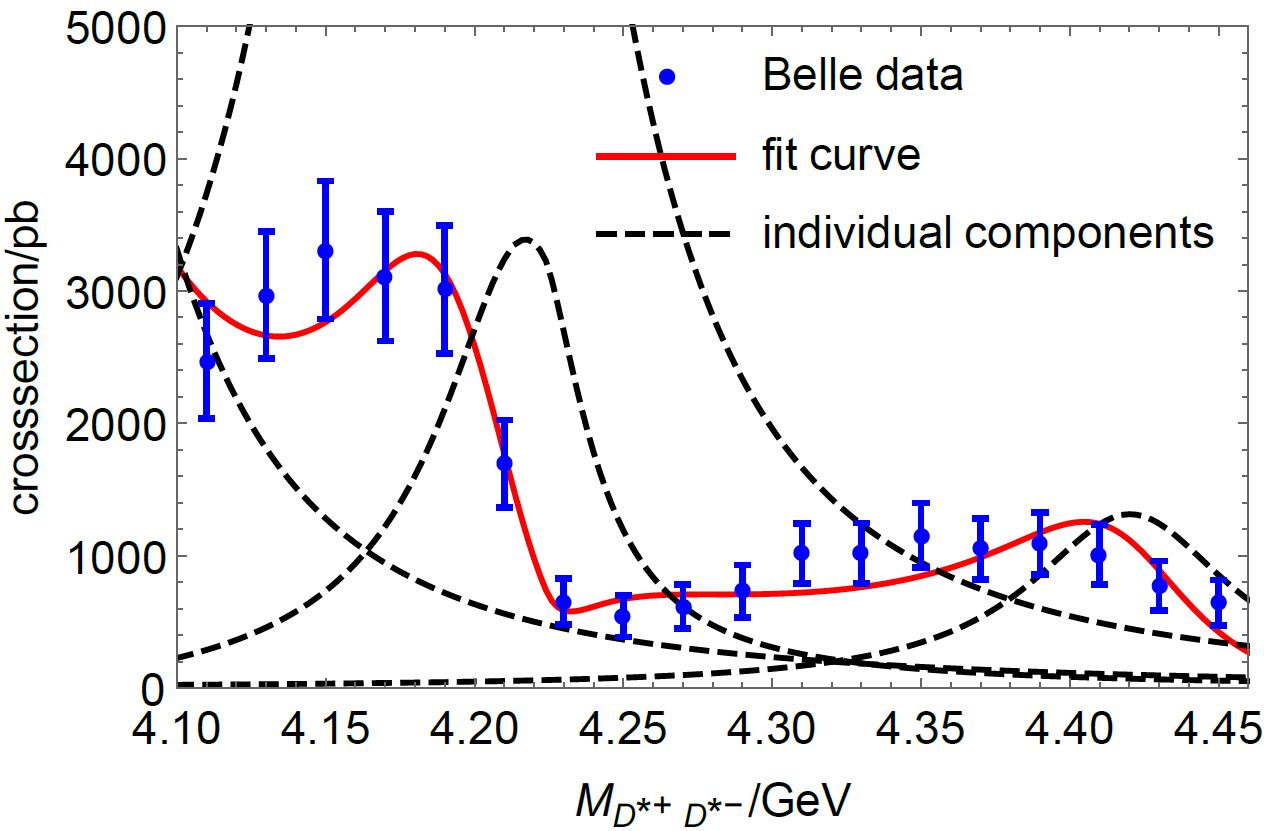}}
\label{fig:dstardstar1}
}
\caption{The results of the fit with $D_s^{*+}D_s^{*-}$ cross section data.The solid curves in all subgrpaphs are the projections from the best fit: (a) and (b) fit to the cross section of $J/\psi \pi^+\pi^-$; (c) fit to $D^0D^{*-}\pi^+$ data, where the dashed black one represents $X(4260)$ components; (d) fit to $h_c\pi^+\pi^-$ data, where the dashed black one indicates $X(4260)$ components; (e) fit to $\psi(2S)\pi^+\pi^-$ data, where the dashed black one describes $X(4260)$ components; (f) fit to $\omega\chi_{c0}$ cross section data, and we do not fit the two data points (orange squres) on the left side since they are below the threshold; (g) fit to $J/\psi\,\eta$ data; (h) fit to $D_s^{*+}D_s^{*-}$ data; (i) fit to $D^+D^{*-}$ cross section, where the dashed curves show the individual components of $\psi(4040)$, $\psi(4160)$, $X(4260)$, respectively; (j) fit to $D^{*+}D^{*-}$ cross section, where the dashed curves show the individual components of $\psi(4040)$, $\psi(4160)$, $X(4260)$ and $\psi(4415)$, respectively. }

\label{fig:fit1}
\end{figure}

For the $J/\psi\pi^+\pi^-$ channel, the experimental data sets come from: BESIII~\cite{Ablikim:2016qzw}, $\sqrt{s}\in[3.81,4.60]$ GeV, 121 data points;
BABAR~\cite{BaBar2012}, $\sqrt{s}\in[4.15,4.47]$ GeV, 17 data points; Belle~\cite{Belley4260}, $\sqrt{s}\in[4.15,4.47]$ GeV, 17 data points.
For the fit to the above data, we adopt the amplitude as Eq.~(14) in {Ref.~\cite{Tang2015}} to describe the $e^+e^-\to\gamma^*\to X(4260)\to J/\psi\pi^+\pi^-$ process and the propagator $\frac{1}{D_X(s)}$\footnote{Here we use $s=q^2$.} of that equation is rewritten as the following form:
\begin{equation}
\frac{1}{D_X(s)}\Rightarrow\frac{1}{s-M_X^2+i\sqrt{s}\Gamma_{tot}}+\frac{c_{11}e^{i\phi_{11}}}{s-M_{4415}^2+i\sqrt{s}\Gamma_{4415}}
+c_{12}e^{i\phi_{12}}e^{-c_{13}(\sqrt{s}-m_{th})}\ \ \mbox{,}
\end{equation}
where $m_{th}$ is the threshold of $J/\psi\pi^+\pi^-$, $M_{4415}$ and $\Gamma_{4415}$ are introduced to represent the mass and width of $\psi(4415)$, respectively;
$\phi_{11}$ and $\phi_{12}$ are interference phases;  $c_{11}$, $c_{12}$, and $c_{13}$ are free constants (see Figures~\ref{fig:jpsipipi1} and \ref{fig:jpsipipibes1} for fit results). %{\color{red}The fit without the BESIII data~\cite{Ablikim:2016qzw} at the energy $\sqrt{s}=3.81$ GeV is also carried out. The result doesn't show anything different and the system error is $0.181\%$. }

\subsubsection{The $D^0D^{*-}\pi^+$, $h_c\pi^+\pi^-$ and $\psi(2S)\pi^+\pi^-$ processes }

The BESIII data for $e^+e^-\rightarrow D^0D^{*-}\pi^+$~\cite{Ablikim:2018vxx} ranging from $\sqrt{s}\in[4.05,4.45]$ GeV are chosen with 51 points.
In $h_c\pi\pi$ channel, the experimental data in $\sqrt{s}\in[4.09,4.32]$ GeV region with 45 data points are also chosen from  BESIII Collaboration~\cite{BESIII:2016adj}. For the $\psi(2S)\pi^+\pi^-$ process, the experimental data come from: BESIII~\cite{Ablikim:2017oaf}, $\sqrt{s}\in[4.085,4.308]$ GeV, 8 data points;
Belle~\cite{Wang:2014hta}, $\sqrt{s}\in[4.11,4.31]$ GeV, 11 data points; BABAR~\cite{Lees:2012pv}, $\sqrt{s}\in[4.13,4.32]$ GeV, 5 data points. The fit formula is
\begin{equation}\label{eq:hcpipi}
\sigma_{i}=\frac{3\pi}{k^2}|\frac{\sqrt{s\Gamma_{ee}\Gamma_{f}}}{s-M_X^2+i\sqrt{s}\Gamma_{tot}}+
\frac{c_{i1}e^{i\phi_{i1}}}{s-M_{4415}^2+i\sqrt{s}\Gamma_{4415}}|^2 \ ,
\end{equation}
where $i=2,3,4$ represents the $D^0D^{*-}\pi^+$, $h_c\pi^+\pi^-$ and $\psi(2S)\pi^+\pi^-$ channels, respectively,
as shown in  Figures~\ref{fig:ddpi1}, ~\ref{fig:hcpipi1}, and~\ref{fig:psi2spipi1}.

\subsubsection{The $\omega\chi_{c0}$ process}

The $\omega\chi_{c0}$ data comes from Ref.~\cite{Ablikim:2015uix}, $\sqrt{s}\in[4.21,4.39]$ GeV, 8 data points and Ref.~\cite{Ablikim:2019apl}, $\sqrt{s}\in[4.199,4.278]$ GeV, 7 data points. The fit formula can be parameterized as
%the Eq.~(\ref{eq:omechi})
\begin{equation}\label{eq:omechi}
\sigma_{\omega\chi_{c0}}=\frac{3\pi}{k^2}|
\frac{\sqrt{s\Gamma_{ee}\Gamma_{\omega\chi_{c0}}}}{s-M_X^2+i\sqrt{s}\Gamma_{tot}}|^2\ ,
\end{equation}
and the fit results are illustrated as Figure~\ref{fig:omegakaic01}.

\subsubsection{The $J/\psi\eta$ process}

Concerning the $J/\psi \eta$ channel, 8 data points ranging from $\sqrt{s}\in[4.09,4.31]$ GeV measured by BESIII~\cite{Ablikim:2015xhk} and 13 data points from $\sqrt{s}\in[4.07,4.31]$ GeV measured by Belle~\cite{Wang:2012bgc} are adopted simultaneously. On account of the influence of $\psi(4160)$, which was taken into account in the fit by Belle in Ref.~\cite{Wang:2012bgc}, the cross section is written as the following:
\begin{equation}\label{eq:jpsieta}
\sigma_{J/\psi\eta}=\frac{3\pi}{k^2}|\frac{\sqrt{s\Gamma_{ee}\Gamma_{J/\psi\eta}}}{s-M_X^2+i\sqrt{s}\Gamma_{tot}}+\frac{c_{51}e^{i\phi_{51}}}{s-M_{4160}^2+i\sqrt{s}\Gamma_{4160}}|^2.
\end{equation}
Besides, $M_{4160}$ and $\Gamma_{4160}$ are introduced to represent the mass and width of $\psi(4160)$, respectively.

\subsubsection{The $D_s^{*+}D_s^{*-}$ process}

In $D_s^{*+}D_s^{*-}$ channel we take the data from Ref.~\cite{Ds*Ds*}, $\sqrt{s}\in[4.23,4.36]$ GeV, 5 data points. Then the cross section reads
%In the fit we add a Breit--Wigner type background term with mass at $4.419$ GeV,
\begin{equation}
\sigma_{D_s^{*+}D_s^{*-}}=\frac{3\pi}{k^2}|\frac{\sqrt{s\Gamma_{ee}\Gamma_{D_s^{*+}D_s^{*-}}}}{s-M_X^2+i\sqrt{s}\Gamma_{tot}}+
\frac{c_{61}e^{i\phi_{61}}}{s-M_{4415}^2+i\sqrt{s}\Gamma_{4415}}|^2.
\end{equation}
See Figure~\ref{fig:dsds1} for fit results.

\subsubsection{The $D^+D^{*-}$ and $D^{*+}D^{*-}$ processes}

For the $D^+D^{*-}$ channel, we take the Belle data~\cite{PhysRevLett.98.092001}, $\sqrt{s}\in[3.93,4.37]$ GeV, with 23 data points, as shown in Figure~\ref{fig:ddstar1}.
Two additional Breit-Wigner resonances and a complex constant are imposed in the fit to simulate the contribution of interference backgrounds:
\begin{equation}\label{eq:ddstar}
\begin{split}
\sigma_{D^+D^{*-}+c.c.}=&\frac{3\pi}{k^2}|\frac{\sqrt{s\Gamma_{ee}\Gamma_{D^+D^{*-}}}}{s-M_X^2+i\sqrt{s}\Gamma_{tot}}+\frac{c_{71}e^{i\phi_{71}}}{s-M_{4040}^2+i\sqrt{s}
\Gamma_{4040}}+\frac{c_{72}e^{i\phi_{72}}}{s-M_{4160}^2+i\sqrt{s}\Gamma_{4160}}\\
&+c_{73}+ic_{74}|^2,
\end{split}
\end{equation}
in which $M_{4040}$ ($\Gamma_{4040}$) is used to describe the mass (width) of $\psi(4040)$.

Additionally, 18 data points released by Belle~\cite{PhysRevLett.98.092001} from $\sqrt{s}\in[4.11,4.45]$ are adopted in the $D^{*+}D^{*-}$ process and the cross section is written as
\begin{equation}\label{eq:dstardstar}
\begin{split}
\sigma_{D^{*+}D^{*-}}=&\frac{3\pi}{k^2}|\frac{\sqrt{s\Gamma_{ee}\Gamma_{D^{*+}D^{*-}}}}{s-M_X^2+i\sqrt{s}\Gamma_{tot}}+\frac{c_{81}e^{i\phi_{81}}}{s-M_{4040}^2+i\sqrt{s}
\Gamma_{4040}}+\frac{c_{82}e^{i\phi_{82}}}{s-M_{4160}^2+i\sqrt{s}\Gamma_{4160}}\\
&+\frac{c_{83}e^{i\phi_{83}}}{s-M_{4415}^2+i\sqrt{s}\Gamma_{4415}}|^2.
\end{split}
\end{equation}
The fit projection is presented in Figure.~\ref{fig:dstardstar1}.

\subsubsection{The fit results}

We have attempted to fit the experimental data with three well established charmonia, $\psi(4040)$, $\psi(4160)$ and $\psi(4415)$, together with other coherent background contributions in the above decay channels. Since $X(4260)$ is our only interest here, the mass of $\psi(4040)$, $\psi(4160)$ and $\psi(4415)$ is fixed whereas the widths are left free in this research. The parameters related to backgrounds in each process are mentioned above, and the widths of $\psi(4040)$, $\psi(4160)$ and $\psi(4415)$ are listed in Table~\ref{tab:coupling1}, which are found in reasonable agreement with the widths given by Particle Data Group~\cite{PhysRevD.98.030001}. The coupling coefficients between $X(4260)$ and different final states are presented in the Table~\ref{tab:coupling1} as well. Especially, with  heavy quark spin symmetry considered, the relationship between $g_{D^+ D^-}$, $g_{D^+D^{*-}+c.c.}$ and $g_{D^{*+}D^{*-}}$ can be calculated~\cite{Voloshin:2012dk} to be $g_{D^+ D^-}:g_{D^+D^{*-}+c.c.}:g_{D^{*+}D^{*-}}=1:4:7$, which is utilized in our fit. Therefore, there is only one parameter in need of describing the coupling coefficient in these three channels. The goodness of the fit is $\chi^2/d.o.f.=292.84/(357-43)=0.93$.
The mass ($M_X$) is determined to be $4.219\pm0.001$ GeV,
and the sum of all partial wave widths excluding the coupling to $D_s^{*+}D_s^{*-}$ is $51.44\pm3.92$ MeV at $\sqrt{s}=M_X$. The value of $g_0$ corresponds to the leptonic decay width $\Gamma_{e^+e^-}=1.314\pm0.066$ keV, which gives a strong support for $X(4260)$ to be a $4^3S_1$ vector charmonium~\cite{Li:2009zu}.

\begin{table}[tbph]
\centering
\caption{
Summary of numerical results associated with the $D_s^{*+}D_s^{*-}$ data.
}
\label{tab:coupling1}
\vskip 0.2cm
\begin{tabular}{cccc}
\hline
\hline
parameters & value \tabularnewline
\hline
$g_0$(MeV) & $23.865\pm0.602$   \tabularnewline

$M_X$(GeV) & $4.219\pm0.001$  \tabularnewline

$\Gamma_0$(GeV) & $0.005\pm0.002$  \tabularnewline

$h_1$ & $-0.0003\pm0.0005$  \tabularnewline

$h_2$ &  $-0.062\pm0.005$ \tabularnewline

$h_3$ & $0.016\pm0.001$ \tabularnewline

$g_{D^0D^{*-}\pi^+}$ & $234.750\pm34.787$    \\

$g_{h_c\pi^+\pi^-}$ &  $68.248\pm22.801$  \\

$g_{\psi(2S)\pi^+\pi^-}$ &  $12.538\pm8.588$  \\

$g_{\omega\chi_{c0}}$ & $0.0007\pm0.0001$  \\

$g_{J/\psi\eta}$(GeV$^{-2}$) &   $0.0002\pm0.00006$  \\

$g_{D_s^{*+}D_s^{*-}}$(GeV$^{-2}$) & $1.101\pm0.004$  \\

$g_{D^+ D^-}$(GeV$^{-2}$) &  $0.004\pm0.0003$  \\

$\Gamma_{4040}$(GeV) & $0.090\pm0.016$  \\

$\Gamma_{4160}$(GeV) & $ 0.080\pm0.020$  \\

$\Gamma_{4415}$(GeV) & $ 0.082\pm0.002$  \\
\hline
\hline
\end{tabular}
\end{table}

The above conclusions are rather stable against variations of background parameterizations. For example, the complex-constant coherent background can be employed in Eq.~(\ref{eq:dstardstar}) for the $D^{*+}D^{*-}$ channel, and the expression is
\label{DD*D*D*}
\begin{equation}
\begin{split}
\sigma_{D^{*+}D^{*-}}=&\frac{3\pi}{k^2}|\frac{\sqrt{s\Gamma_{ee}\Gamma_{D^{*+}D^{*-}}}}{s-M_X^2+i\sqrt{s}\Gamma_{tot}}+\frac{c_{81}e^{i\phi_{81}}}{s-M_{4040}^2+i\sqrt{s}
\Gamma_{4040}}+\frac{c_{82}e^{i\phi_{82}}}{s-M_{4160}^2+i\sqrt{s}\Gamma_{4160}}\\
&+\frac{c_{83}e^{i\phi_{83}}}{s-M_{4415}^2+i\sqrt{s}\Gamma_{4415}}+c_{84}+i c_{85}|^2.
\end{split}
\end{equation}
It turns out that the fit quality is $\chi^2/d.o.f.=282.38/(357-45)=0.91$ and the fit results for each decay channel make practically little difference.

\subsection{Discussions on the fit without $D_s^{*+}D_s^{*-}$ data}\label{fitwithoutdsds}

\begin{figure}[htbp]
\center
\subfigure[$J/\psi \pi^+\pi^-$]{
\scalebox{1.2}[1.2]{\includegraphics[width=0.25\textwidth]{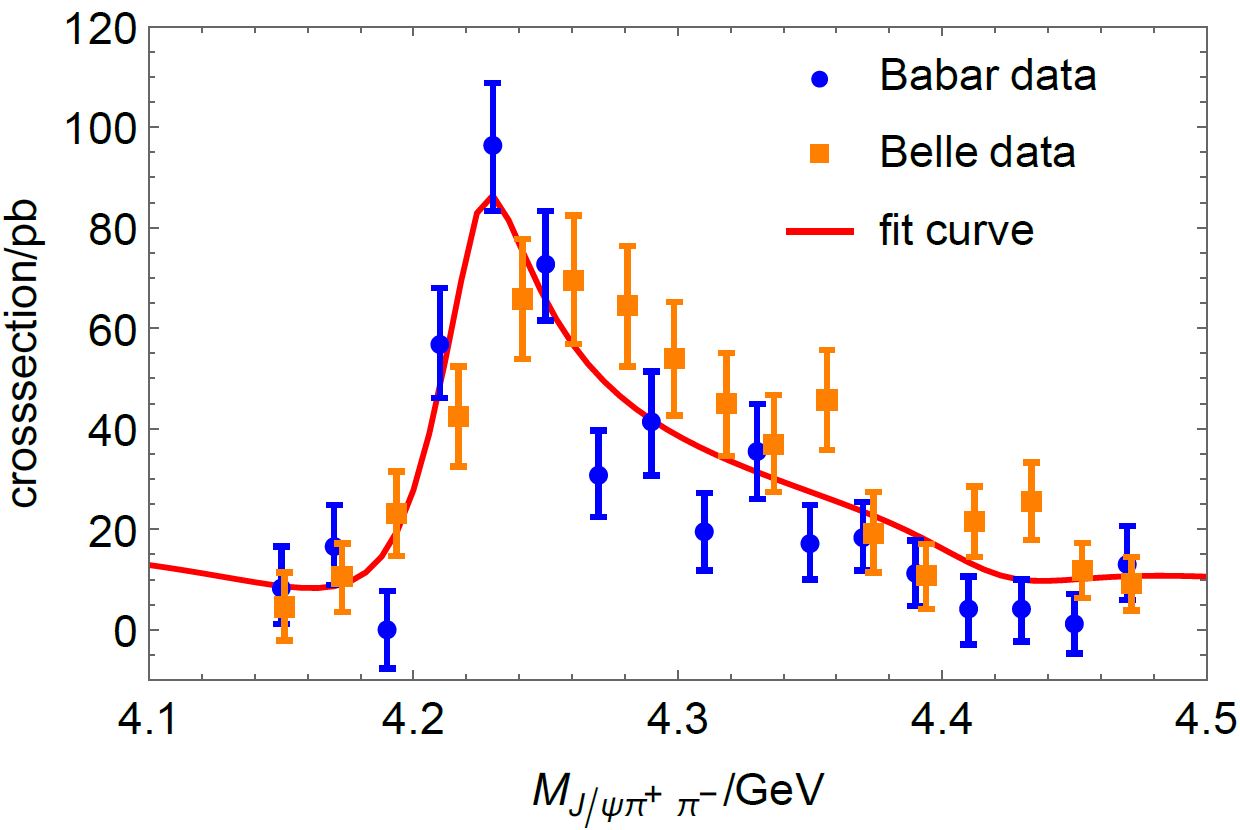}}
\label{fig:jpsipipi2}
}
\subfigure[$J/\psi \pi^+\pi^-$]{
\scalebox{1.2}[1.2]{\includegraphics[width=0.25\textwidth]{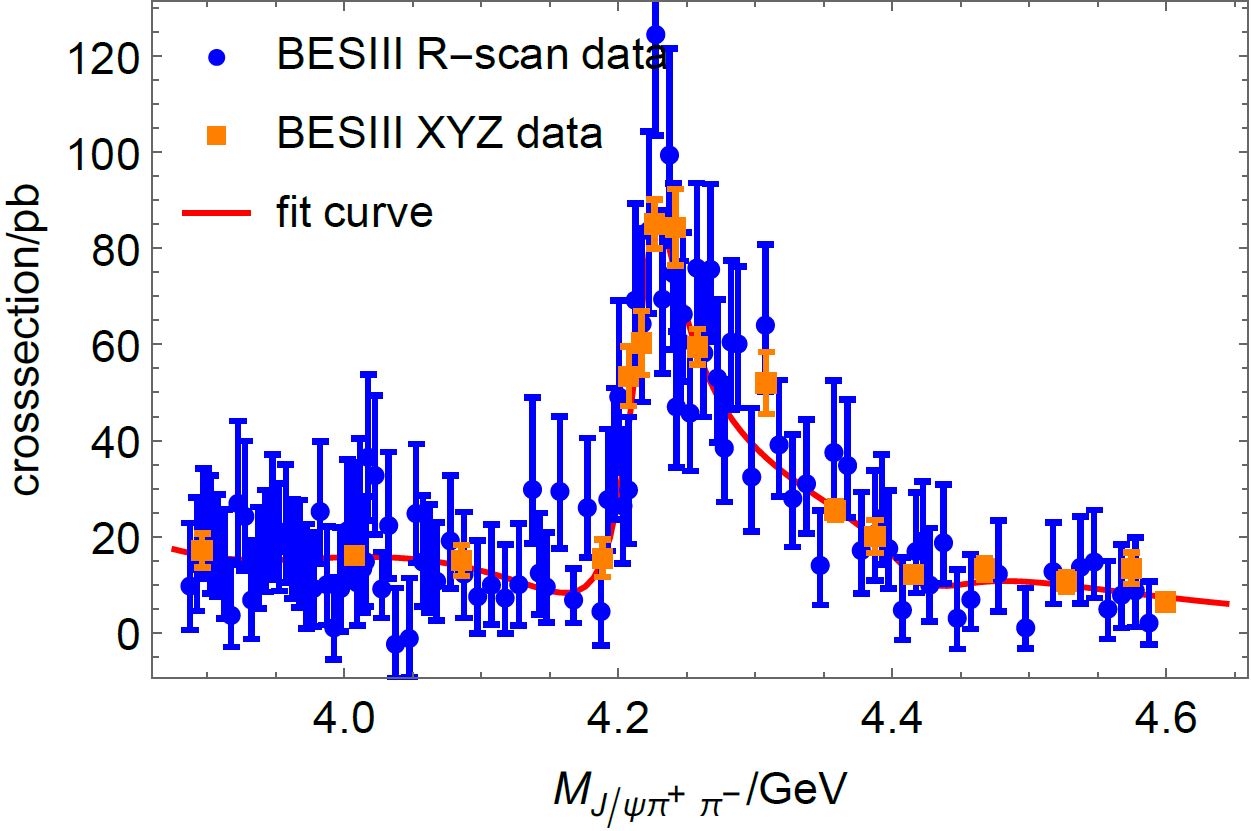}}
\label{fig:jpsipipibes2}
}
\subfigure[$D^0D^{*-}\pi^+$]{
\scalebox{1.2}[1.2]{\includegraphics[width=0.25\textwidth]{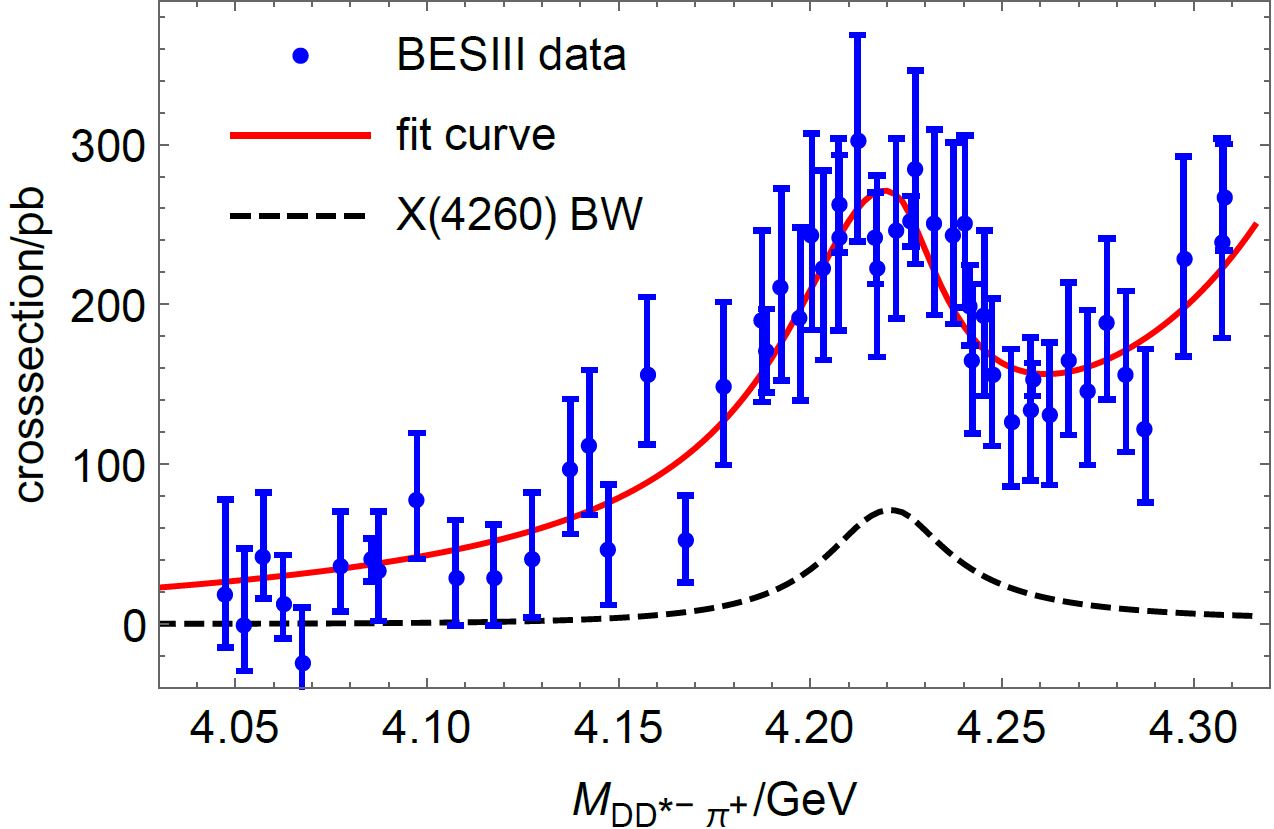}}
\label{fig:ddpi2}
}
\subfigure[$h_c\pi^+\pi^-$]{
\scalebox{1.2}[1.2]{\includegraphics[width=0.25\textwidth]{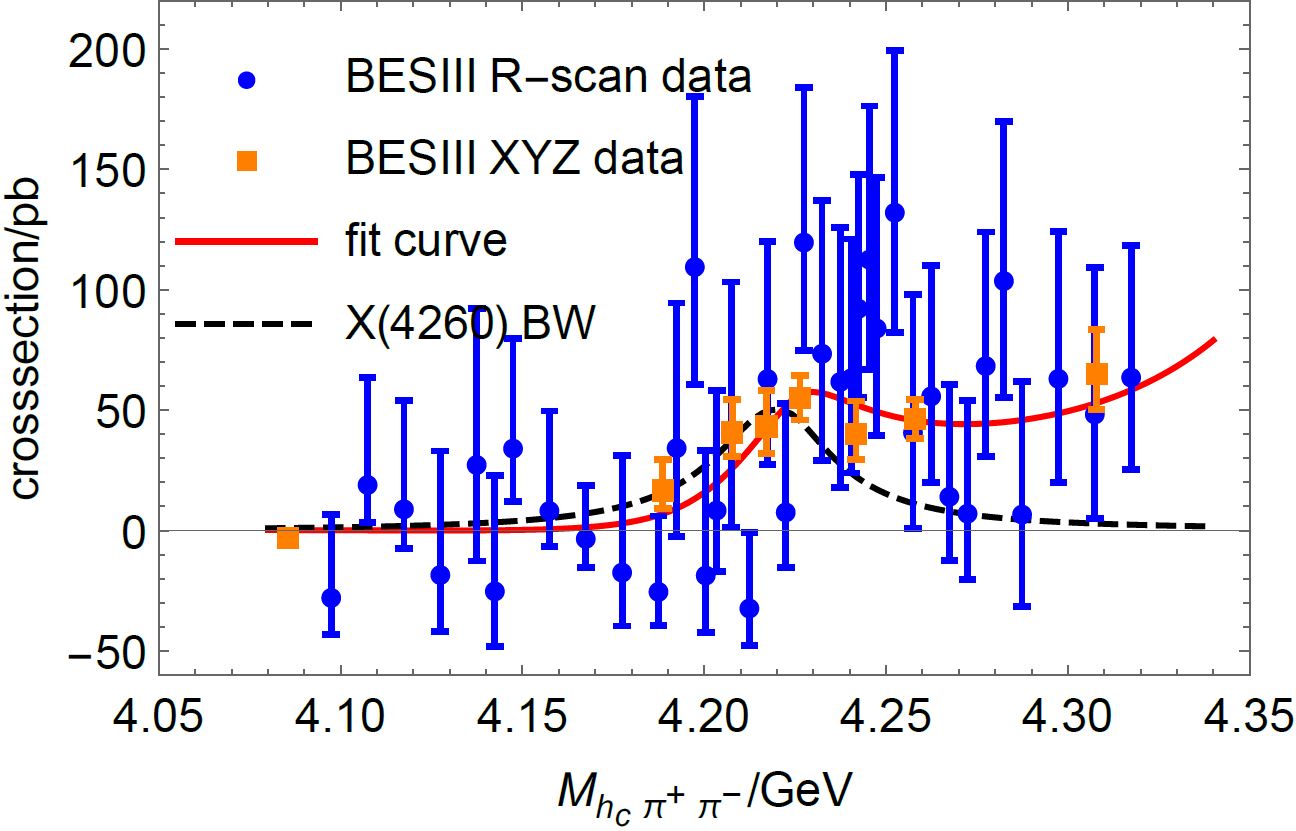}}
\label{fig:hcpipi2}
}
\subfigure[$\psi(2S)\pi^+\pi^-$]{
\scalebox{1.2}[1.2]{\includegraphics[width=0.25\textwidth]{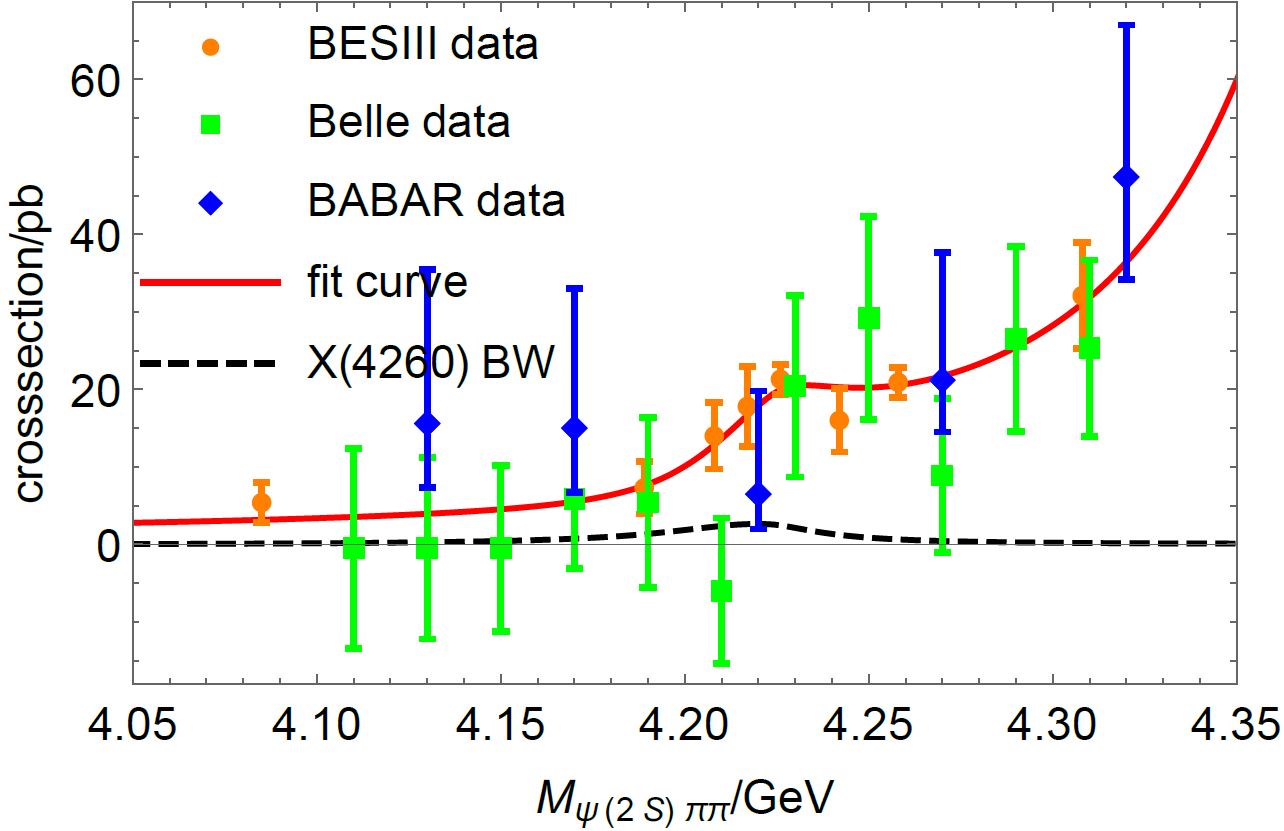}}
\label{fig:psi2spipi2}
}
\subfigure[$\omega\chi_{c0}$]{
\scalebox{1.2}[1.2]{\includegraphics[width=0.25\textwidth]{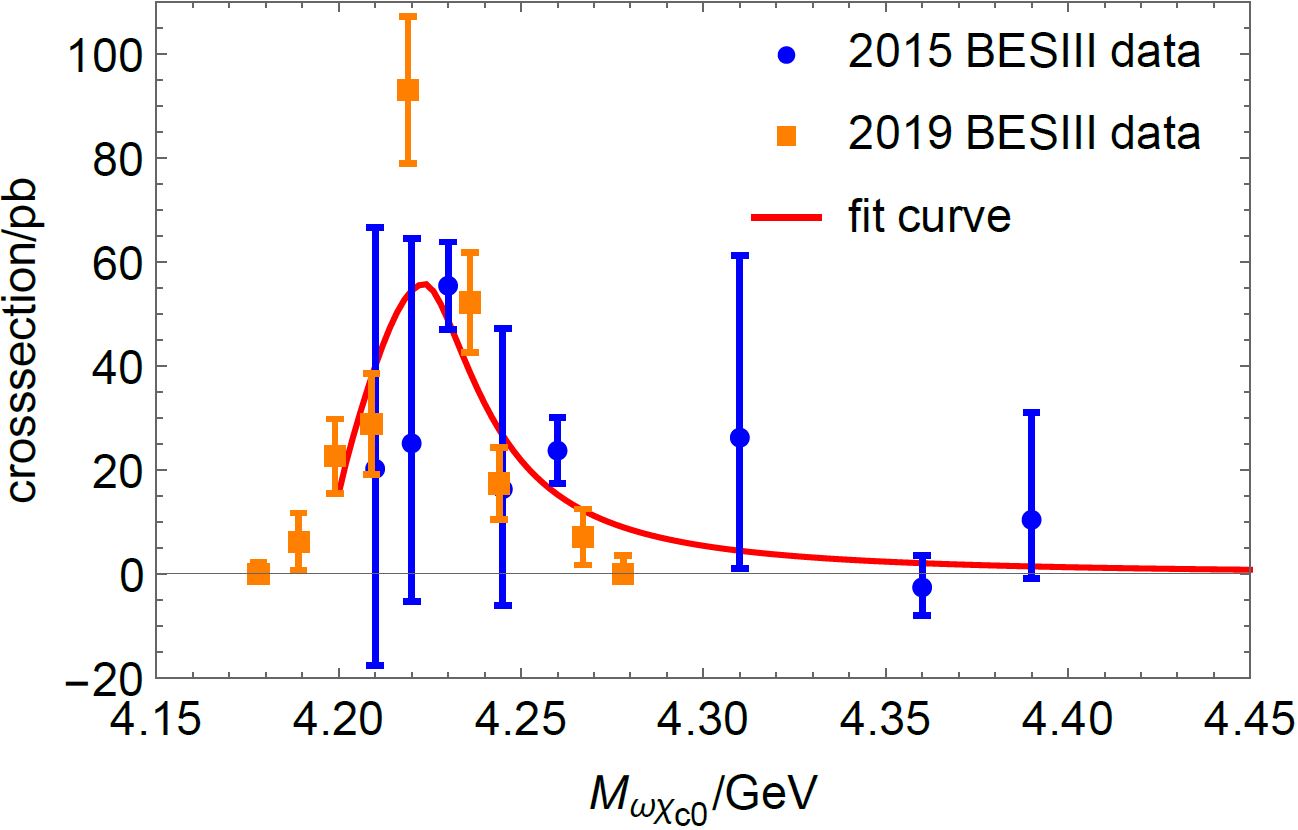}}
\label{fig:omegakaic02}
}
\subfigure[$J/\psi\,\eta$ ]{
\scalebox{1.2}[1.2]{\includegraphics[width=0.25\textwidth]{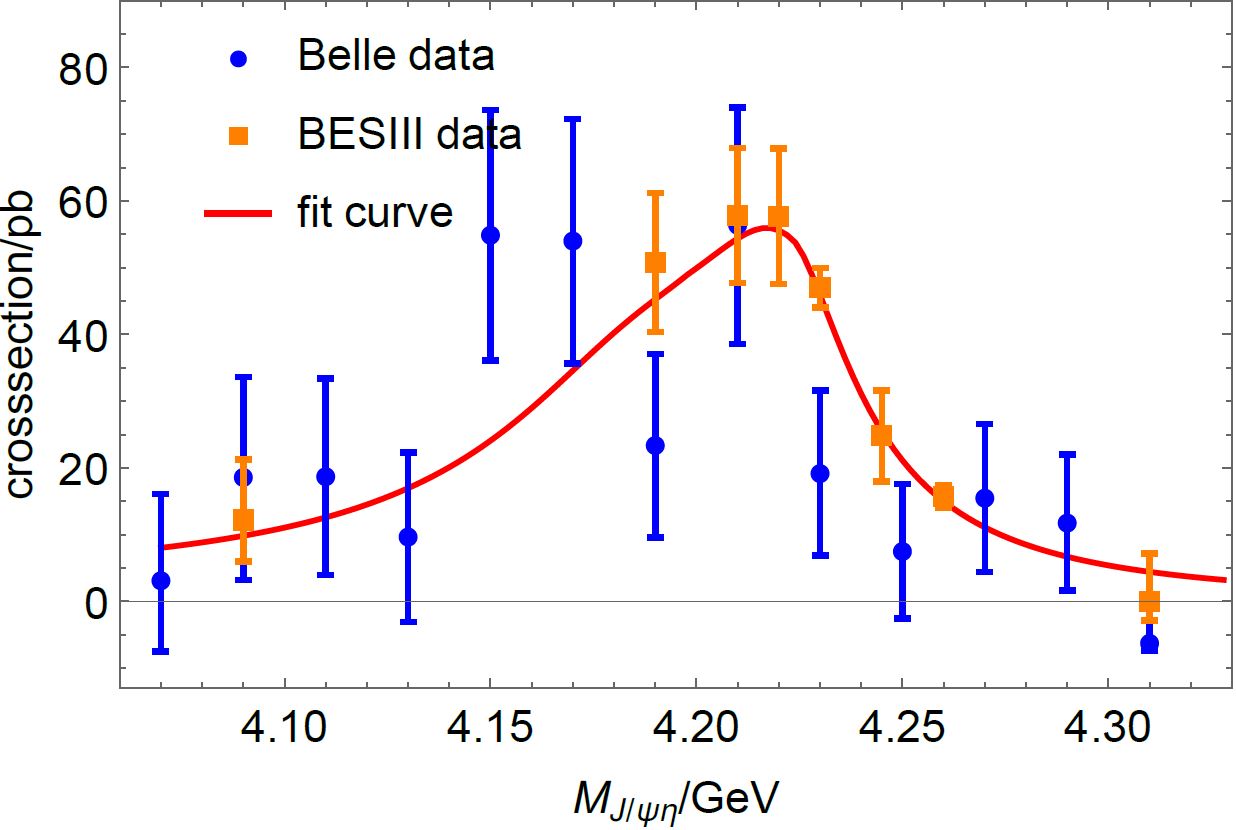}}
\label{fig:jpsieta2}
}
\subfigure[$D^+D^{*-}$]{
\scalebox{1.2}[1.2]{\includegraphics[width=0.25\textwidth]{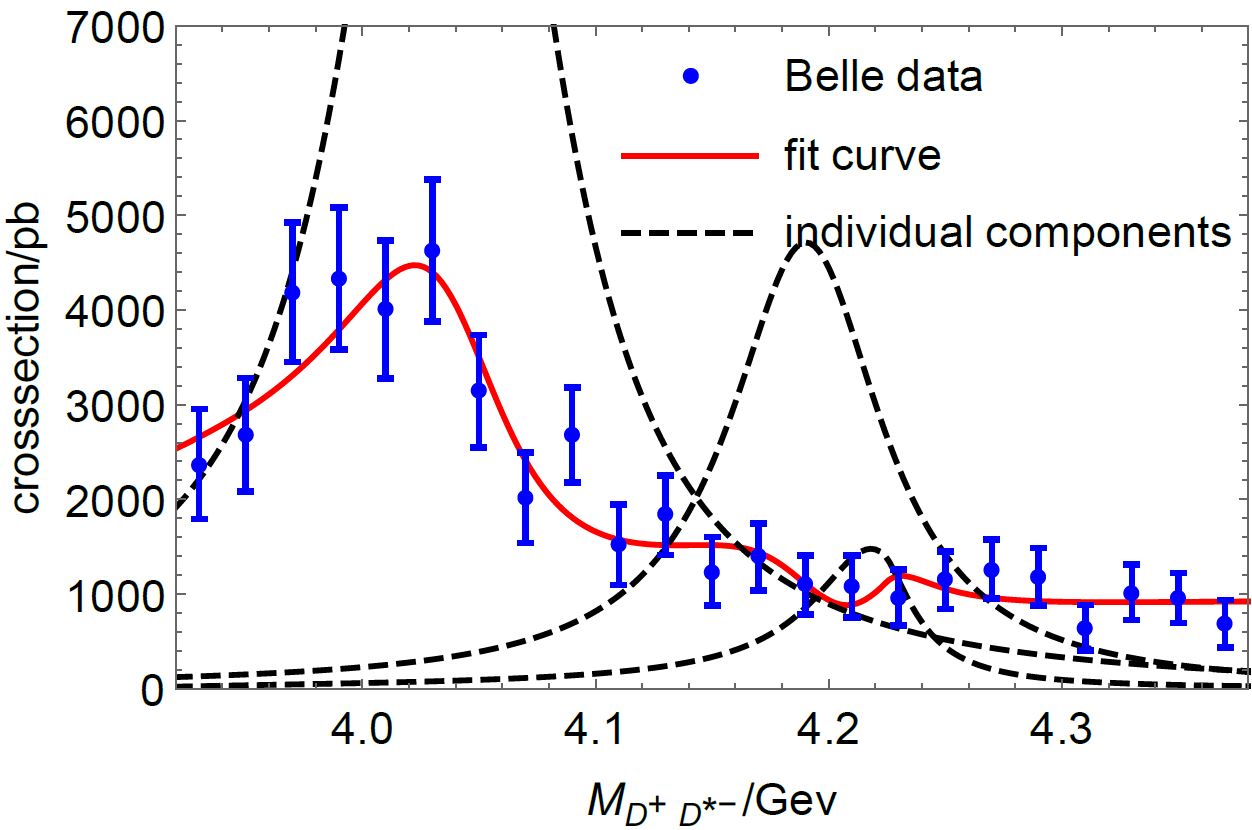}}
\label{fig:ddstar2}
}
\subfigure[$D^{*+}D^{*-}$]{
\scalebox{1.2}[1.2]{\includegraphics[width=0.25\textwidth]{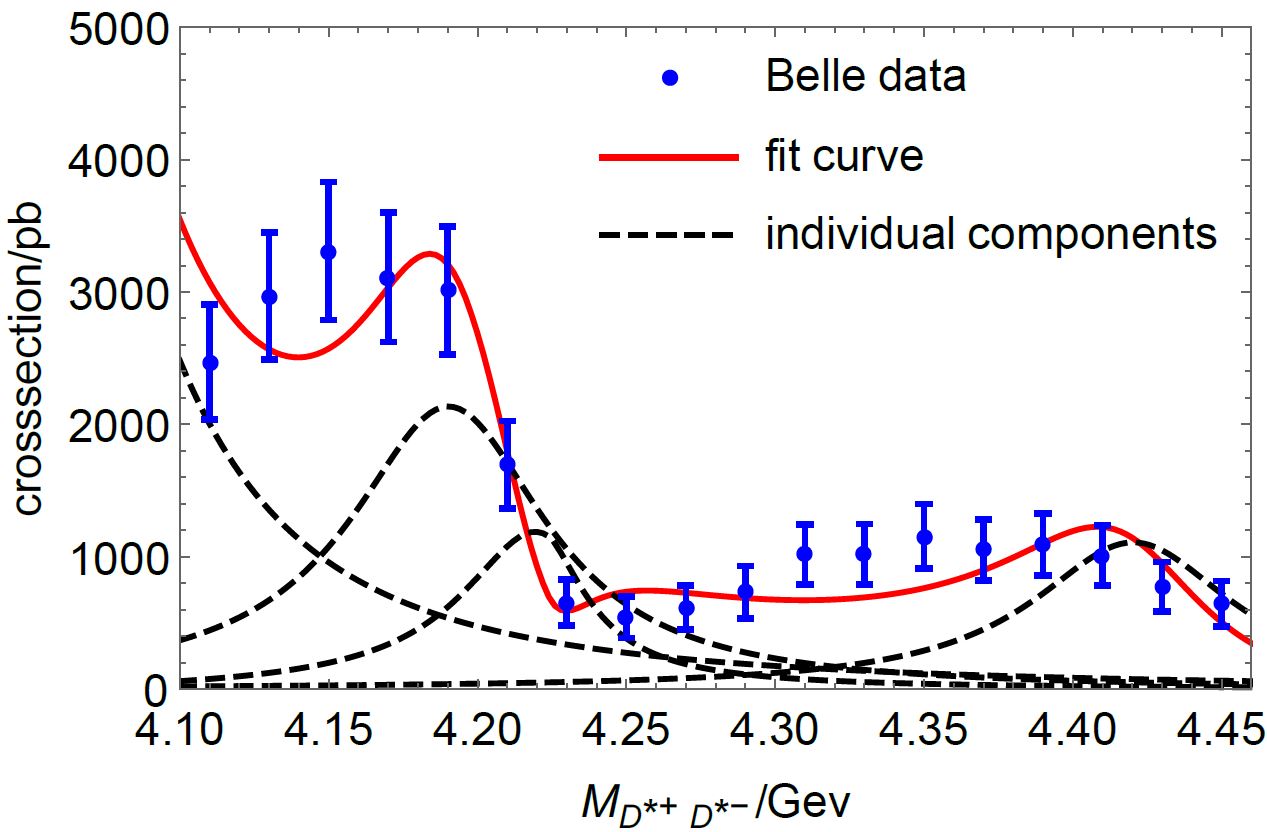}}
\label{fig:dstardstar2}
}
\caption{The results of the fit without $D_s^{*+}D_s^{*-}$ cross section data, where the descriptions of the components in subgraphs are similar to those of Fig.~\ref{fig:fit1}}

\label{fig:fit2}
\end{figure}

Since the $D_s^{*+}D_s^{*-}$ cross section data from BESIII~\cite{Ds*Ds*} are preliminary, the program without fitting the $D_s^{*+}D_s^{*-}$ has also been carried out. In this subsection, the total width of the $X(4260)$ propagator is also Eq.~(\ref{eq:gammatotal}), which includes the $D_s^{*+}D_s^{*-}$ decay width, even though the data are not fitted. It is noticed that the branching ratios of each decay channel remain similar whether the program includes fitting the $D_s^{*+}D_s^{*-}$ cross section data or not. Likewise, the fit formula for the $J/\psi\pi^+\pi^-$, $D^0D^{*-}\pi^+$, $h_c\pi^+\pi^-$, $\psi(2S)\pi^+\pi^-$, $\omega\chi_{c0}$, $J/\psi\eta$, $D^+D^{*-}$ and $D^{*+}D^{*-}$ processes are used as the forms in the section~\ref{fitwithdsds} respectively. The fit results are displayed in Figure~\ref{fig:fit2}.

The widths of $\psi(4040)$, $\psi(4160)$ and $\psi(4415)$ as well as the coupling coefficients between $X(4260)$ and different final states are presented in Table~\ref{tab:coupling2}. The fit quality is $\chi^2/d.o.f.=291.86/(352-41)=0.94$. The value of $g_0$ corresponds to a leptonic decay width $\Gamma_{e^+e^-}=0.385\pm0.060$ keV, which may imply that $X(4260)$ is a mixture of $4^3S_1$ and $3^3D_1$ $c\bar c$ charmonium state~\cite{Li:2009zu}.

\begin{table}[tbph]
\centering
\caption{
Summary of relevant parameters without the $D_s^{*+}D_s^{*-}$ data. }\label{tab:coupling2}
\vskip 0.2cm
\begin{tabular}{cc}
\hline
\hline
parameters & value \tabularnewline
\hline
$g_0$(MeV) & $12.928\pm1.000$  \tabularnewline

$M_X$(GeV) & $4.221\pm0.001$  \tabularnewline

$\Gamma_0$(GeV) & $0.000\pm0.0002$  \tabularnewline

$h_1$ & $0.0004\pm0.0008$  \tabularnewline

$h_2$ &  $0.103\pm0.008$ \tabularnewline

$h_3$ & $-0.026\pm0.002$ \tabularnewline

$g_{D^0D^{*-}\pi^+}$ & $603.980\pm122.580$    \\

$g_{h_c\pi^+\pi^-}$ &  $197.150\pm71.981$  \\

$g_{\psi(2S)\pi^+\pi^-}$ &  $31.475\pm21.230$  \\

$g_{\omega\chi_{c0}}$ & $0.002\pm0.0003$  \\

$g_{J/\psi\eta}$(GeV$^{-2}$)  &   $0.0003\pm0.0001$ \\

$g_{D_s^{*+}D_s^{*-}}$(GeV$^{-2}$) & $1.112\pm0.002$  \\

$g_{D^+ D^-}$(GeV$^{-2}$) &  $0.004\pm0.0002$ \\

$\Gamma_{4040}$(GeV) & $0.090\pm0.014$  \\

$\Gamma_{4160}$(GeV) & $ 0.080\pm0.014$  \\

$\Gamma_{4415}$(GeV) & $ 0.082\pm0.004$ \\
\hline
\hline
\end{tabular}
\end{table}

In addition, another solution to the fit without $D_s^{*+}D_s^{*-}$ cross section data is found, with the fit quality to be $\chi^2/d.o.f.=288.41/(352-41)=0.93$, where $g_0=21.212\pm1.604$ MeV.
The leptonic width is changed to $1.038\pm0.157$ keV, which is different from the previous solution but is similar to the fit with $D_s^{*+}D_s^{*-}$ cross section data. We find that, unfortunately, the destructive interference in the open charm channels is rather unstable, which makes it impossible to distinguish the physical one between the two solutions with the similar fit quality, and we can not figure out another solution better than the two. So only the range of the leptonic width rather than a definitive value may be trustworthy. Even so, the leptonic width of over hundreds eV still favor $X(4260)$ as a conventional charmonium.

\begin{figure}
\centering
\includegraphics[width=80mm]{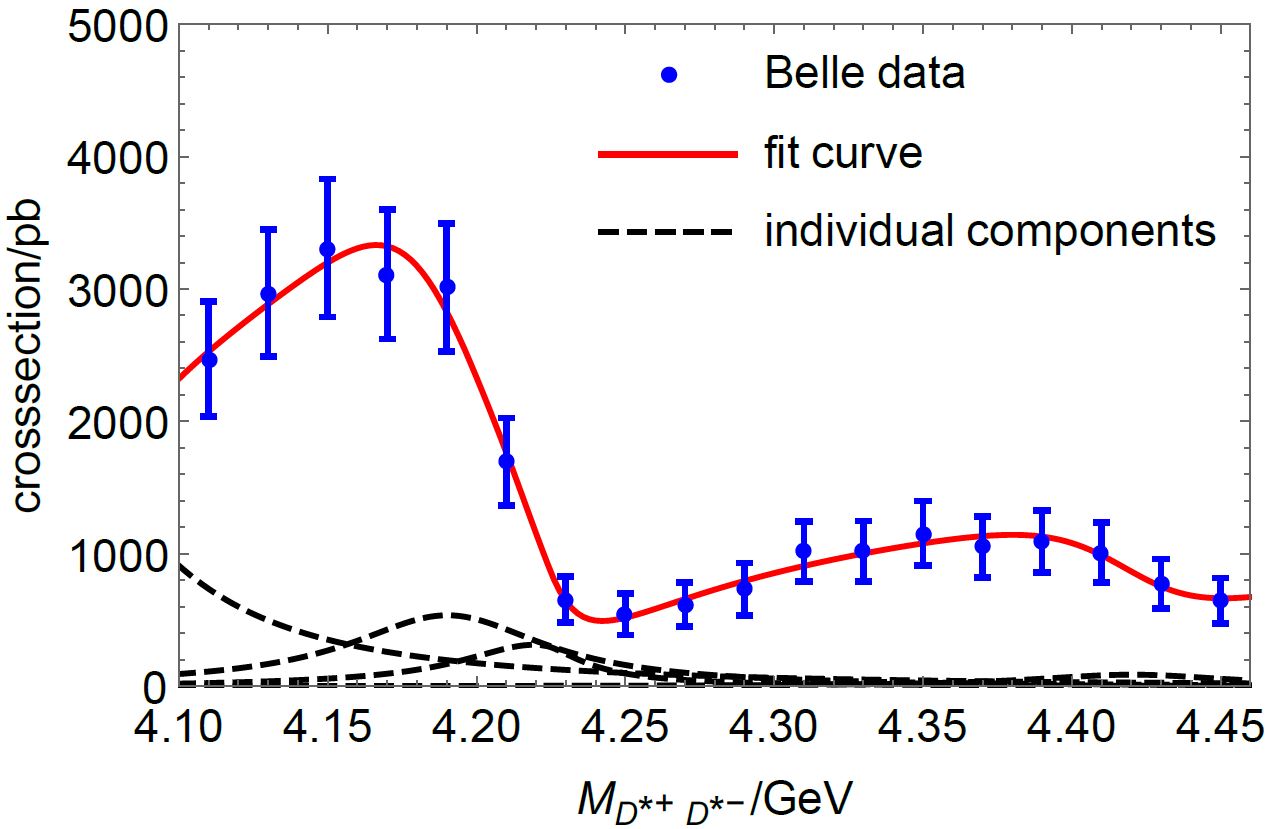}
  \caption{Given the complex coherent background in $D^{*+}D^{*-}$ process, the fit result of the $D^{*+}D^{*-}$ cross section. The blue dots come from Belle results~\cite{PhysRevLett.98.092001}. The solid curve is the projection from the best fit. The dashed curves show
the individual components of $\psi(4040)$, $\psi(4160)$, $X(4260)$ and $\psi(4415)$, respectively. The contribution of the complex background is not presented.  }\label{fig:dstardstar3}
 \end{figure}

Owing to the instability brought by the open charm channels, then we add a complex coherent background in $D^{*+}D^{*-}$ channel following the strategy of section~\ref{DD*D*D*} to test the stability of outputs against the variation of backgrounds. The fit is plotted in Figure~\ref{fig:dstardstar3}. The fit quality is $\chi^2/d.o.f.=275.67/(352-43)=0.89$. It is found, however, unlike the result of section~\ref{DD*D*D*}, the fit is not quite stable here. The difference is clearly seen when comparing Figure~\ref{fig:dstardstar1} and Figure~\ref{fig:dstardstar3}: the interference between different resonances are done in rather different manner. The leptonic width behaves quite differently, with a value of $\Gamma_{e^+e^-}=0.317\pm0.064$ keV, comparing with the result of $D^{*+}D^{*-}$ channel without constant background.

Moreover, we also find another new solution to the fit with a complex coherent background in $D^{*+}D^{*-}$ channel. The fit quality is $\chi^2/d.o.f.=277.20/(352-43)=0.90$ and $\Gamma_{e^+e^-}=1.077\pm0.164$ keV, which is compatible with the second solution to the fit without a complex coherent background in $D^{*+}D^{*-}$ channel. The physical solution is not able to be picked out from the two solutions.

\subsection{Summary and discussions on numerical fits}
To compare with the fits discussed above and to further test the stability of the whole fit program, here we also test the fit without including the $D^{+}D^{*-}$ and $D^{*+}D^{*-}$ cross section, but with the $D_s^{*+}D_s^{*-}$ data included. The fit quality is $\chi^2/d.o.f.=266.82/(316-29)=0.93$, with the leptonic width $1.119\pm0.081$ keV. Besides, the constant width $\Gamma_{0}$ describing other decay channels is raised up to $48.95$ MeV, which indicates that the dominant decays are out of $J/\psi\pi^+\pi^-$, $h_c\pi^+\pi^-$, $D^0D^{*-}\pi^+$, $\psi(2S)\pi^+\pi^-$, $\omega\chi_{c0}$,  $J/\psi\eta$ and $D_s^{*+}D_s^{*-}$. As is mentioned above, the decay widths of $D^{+}D^{-}$, $D^{+}D^{*-}$ and $D^{*+}D^{*-}$ are indeed about $40$ MeV, which implies that the fit programs are self-consistent.

For the fit  excluding both  $D^+D^{*-}$, $D^{*+}D^{*-}$ and $D_s^{*+}D_s^{*-}$ cross sections, the parameter $g_0$ is $4.58$ MeV with $\Gamma_{e^+e^-}=0.048\pm0.018$ keV, which approches the result given in Ref.~\cite{Tang2015}. However, we believe this scenario does not have much chance to be physically correct, since there is no reason $a$ $priori$ to exclude the couplings between $X(4260)$ and these states. We may conclude, in the most conservative situation, one still get a leptonic width well above $10^2$ eV, which is compatible with the upper limit 580 eV~\cite{Mo:2006ss} obtained by reanalyzing BESII R-scan data~\cite{Bai:1999pk,Bai:2001ct}. If taking the $D_s^{*+}D_s^{*-}$ data into account, the leptonic width will easily exceed 1 keV.

In addition, the branching ratio of $X(4260)\to J/\psi \pi^+ \pi^-$ is determined to be $\mathcal{O} (10^{-4})$, so that the product $\Gamma(X(4260)\to e^+e^-)\times\mathcal{B} (X(4260)\to J/\psi \pi^+ \pi^-)$ is about $\mathcal{O}(10^{-1})$ eV, which is smaller than the old data~\cite{Babary4260} measured by BABAR Collaboration in 2005. Yet the product $\Gamma(X(4260)\to e^+e^-)\times\mathcal{B} (X(4260)\to J/\psi \pi^+ \pi^-)$ is the same order of magnitude as those of other charmonia~\cite{PhysRevD.98.030001}(such as $\psi(4040)$ and $\psi(4160)$) to $J/\psi \pi^+ \pi^-$ near 4 GeV. It may further imply that $X(4260)$ serves as a conventional charmonium state.

The pole location of $X(4260)$ propagator with $D_s^{*+}D_s^{*-}$ coupling is also searched for in order to achieve a better understanding on the nature of $X(4260)$. Since $D\bar D$, $D \bar D^{*}$ and $D^{*} \bar D^{*}$ channels play a vital role and the threshold of $D_s^{*+}D_s^{*-}$ channel is close to the location of $X(4260)$, the complicated multi-sheets structure of the complex plane is simplified as 4 sheets defined in Table~\ref{tab:riemannsheet}. It is noticed that there are two  poles on sheet II and III as shown in Table~\ref{tab:pole}.
The pole width is a bit smaller than the line shape width excluding the coupling to $D_s^{*+}D_s^{*-}$. We think this is well understood and be a typical situation in $p$ waves, when the pole lies below the second threshold. Since one partial width is $\propto k^3$ where $k$ is the 2nd channel momentum,  below the second threshold the $k^2$ factor provides an additional minus sign. However, it should be emphasized that the appearance of two poles, is not a manifestation of the ``elementariness" of $X(4260)$, since the pole counting criteria only applies when couplings are in $s$ waves.~\footnote{The pole counting criteria is originally proposed in Ref.~\cite{morgan}, and has been applied to  the discussions of the $X,Y,Z$ states in, for example, Refs.~\cite{ZcGong, Cao:2019wwt}--\cite{Zhang:2009bv}.}

\begin{table}[tbph]
\centering
\caption{Definition of the four Riemann sheets with $D_s^{*+}D_s^{*-}$ channel and the $D\bar D$, $D \bar D^{*}$ and $D^{*} \bar D^{*}$ channels as a whole.}
\begin{tabular}{ccccc}
\noalign{\smallskip}\hline
\hline
  & sheet I & sheet II & sheet III &sheet IV \tabularnewline
\hline\noalign{\smallskip}
$D\bar D + D\bar D^{*} + D^{*} \bar D^{*}$ &  $+$ & $-$ & $-$& $+$   \tabularnewline
$D_s^{*+}D_s^{*-}$  &  $+$ & $+$ & $-$& $-$ \tabularnewline
\noalign{\smallskip}\hline
\hline
\end{tabular}
\label{tab:riemannsheet}
\end{table}

\begin{table}[tbph]
\centering
\caption{Pole positions of $X(4260)$. The value of $\sqrt{s}\equiv M_{\rm pole}-i\Gamma_{\rm pole}/2$ is given in unit of GeV.}
\begin{tabular}{cccc}
\noalign{\smallskip}\hline
\hline
sheet I & sheet II & sheet III &sheet IV \tabularnewline
\hline
 $-$ &  $4.218-i0.015$ & $4.218-i0.009$& $-$ \tabularnewline
\hline
\hline
\end{tabular}
\label{tab:pole}
\end{table}

\section{Conclusions}\label{conclusion}

Studies on $X(4260)$ resonance play an important role in deepening our understandings on exotic particles and strong interactions. Ref.~\cite{Tang2015} pointed out  that $X(4260)$ is a confining state with a very small leptonic decay width which is hard to be understood by  a simple quark model calculation. Thanks to the new experimental data available, a correct understanding gradually emerges, as we believe: a combined fit with the ``old" $D^+D^{*-}$ and $D^{*+}D^{*-}$ data -- even though  there is no apparent  $X(4260)$ peak showing up in these channels -- reveals that the $X(4260)$ can have a sizable leptonic width up to at least $\mathcal{O}(10^2)$eV. Further the fit including the $D_s^{*+}D_s^{*-}$ data can raise the value up to 1 keV. It is worth mentioning that Ref.~\cite{BESIII:2020peo} gives the muonic width to be from 1.09 to 1.53 keV in the range from 4212.8 to 4219.4 MeV, which provides a strong support to our results.\footnote{We point out that the first version of the present manuscript  was put on arXiv 4 months before Ref.~\cite{BESIII:2020peo}.} In Ref.~\cite{Li:2009zu}, which uses a screening potential instead of a linear confining potential to calculate the spectrum, it is estimated that a $4 ^3S_1$ state has a leptonic width $\sim 1$ keV, whereas a $3^3D_1$ state has a leptonic width $\sim$ 50 eV. Hence the smaller $\Gamma_{e^+e^-}$ ($\sim300$ eV) obtained in this paper may be provided by a $3 ^3D_1$ state (maybe a small portion of $2^3D_1$ state as well) mixed with certain portion of $4^3S_1$ state, and the larger value estimated in this paper may corresponds to a $4 ^3S_1$ state, and is probably largely renormalized by the $D_s^{*+}D_s^{*-}$ continuum.  To further determine  the accurate portion of these mixing is still an open question awaiting more fine studies both theoretically and experimentally.

\section{Acknowledgement}
We are grateful to illuminating discussion with Kuang-ta Chao, Ce Meng and Chang-Zheng~Yuan at early stage of this work. This work is supported in part by National Natural Science Foundations of China (NSFC) under Contract Nos. 10925522, 11021092;
and China Postdoctoral Science Foundation under Contract Number 2020M680500.

\renewcommand\refname{Reference}


\begin{thebibliography}{99}
%\bibliographystyle{h-physrev}
\bibitem{Babary4260}
   B.~Aubert {\it et al.} (BABAR Collaboration),
   Phys.\ Rev.\ Lett.\ {\bf 95}, 142001 (2005).
\bibitem{CLEOy4260a}
   T.~E.~Coan {\it et al.} (CLEO Collaboration),
   Phys.\ Rev.\ Lett.\ {\bf 96}, 162003 (2006).
\bibitem{Belley4260}
   C.~Z.~Yuan {\it et al.} (Belle Collaboration),
   Phys.\ Rev.\ Lett.\ {\bf 99}, 182004 (2007).

\bibitem{PhysRevD.98.030001}
   M.~Tanabashi {\it et al.} (Particle Data Group),
   Phys.\ Rev.\ D {\bf 98}, 030001 (2018).

\bibitem{BaBar2012}
   J.~P.~Lees {\it et al.} (BABAR Collaboration),
   Phys.\ Rev.\ D {\bf 86}, 051102 (2012).
%5



\bibitem{nqm}
   S.~Godfrey and N.~Isgur,
   Phys.\ Rev.\ D {\bf 32}, 189 (1985).
\bibitem{review}
   E.~S.~Swanson,
   Phys.\ Rept.\ {\bf 429}, 243 (2006).
\bibitem{Zhou:2013ada} Z.~Y.~Zhou,  Z.~G.~Xiao, Eur. Phys. J. {\bf A}50 (2014)165.

\bibitem{Zhou:2011sp}
    Z.~Y.~Zhou, Z.~G~Xiao, Phys. Rev. {\bf D}84 (2011)034023.
%9
 \bibitem{Xue:2017xpu}
   S.~R.~Xue, H.~J.~Jing, F.~K.~ Guo, and Q.~Zhao,
   Phys.\ Lett.\ B {\bf 779} (2018) 402-408.
\bibitem{Qin:2016spb}
   W.~Qin, S.~R.~Xue, and Q.~Zhao,
   Phys.\ Rev.\ D {\bf 94}, 054035 (2016).
\bibitem{Ding:2008gr}
   G.~J.~Ding,
   Phys.\ Rev.\ D {\bf 79}, 014001 (2009).
\bibitem{Close:2009ag}
   F.~Close, and C.~Downum,
   Phys.\ Rev.\ Lett.\ {\bf 102}, 242003 (2009).
%13
\bibitem{Li:2013bca}
   M.~T.~Li, W.~L.~Wang, Y.~B.~Dong and Z.~Y.~Zhang,
   arXiv:1303.4140.

\bibitem{YDChen2013}Y.~D.~Chen, C.~F.~Qiao, P.~N.~Shen,  Z.~Q.~Zeng, Phys.\  Rev.\ D {\bf 88}, 114007 (2013)

\bibitem{Wang:2019mhs}
   J.~Z.~Wang, D.~Y.~Chen, X.~Liu and T.~Matsuki,
   Phys.\ Rev.\ D {\bf 99}, 114003 (2019).
\bibitem{Huang:2019agb}
   Q.~Huang, D.~Y.~Chen, X.~Liu and T.~Matsuki,
   Eur.\ Phys.\ J.\ C {\bf 79}, 613 (2019).
\bibitem{Chen:2017uof}
   D.~Y.~Chen, X.~Liu and T.~Matsuki,
   Eur.\ Phys.\ J.\ C {\bf 78}, 136 (2018).
\bibitem{Li:2009zu}
   B.~Q.~Li and K.~T.~Chao,
   Phys.\ Rev.\ D {\bf 79}, 094004 (2009).
\bibitem{LlanesEstrada:2005hz}
   F.~J.~Llanes-Estrada,
   Phys.\ Rev.\ D {\bf 72}, 031503 (2005).
\bibitem{Kou:2005gt}
   E.~Kou and O.~Pene,
   Phys.\ Lett.\ B {\bf 631} (2005) 164-169.
\bibitem{Zhu:2005hp}
   S.~L.~Zhu,
   Phys.\ Lett.\ B {\bf 625} (2005) 212.


\bibitem{Coito:2019cts}
   S.~Coito, and F.~Giacosa,
   Acta Phys.\ Polon.\ B. {\bf 51}, no.8, 1713-1737 (2020)

\bibitem{PhysRevD.77.011103}
   G.~Pakhlova {\it et al.} (Belle Collaboration),
   Phys.\ Rev.\ D {\bf 77}, 011103 (2008).
\bibitem{PhysRevLett.98.092001}
   G.~Pakhlova {\it et al.} (Belle Collaboration),
   Phys.\ Rev.\ Lett.\ {\bf 98}, 092001 (2007).
\bibitem{Babar2009}
   B.~Aubert {\it et al.} (BABAR Collaboration),
   Phys.\ Rev.\ D {\bf 79}, 092001 (2009).
\bibitem{Tang2015}
   L.~Y.~Dai, M.~Shi, G.~Y.~Tang, and H.~Q.~Zheng,
   Phys.\ Rev.\ D {\bf 92}, 014020 (2015).
\bibitem{Ablikim:2014qwy}
   M.~Ablikim {\it et al.} (BESIII Collaboration),
   Phys.\ Rev.\ Lett.\ {\bf 114}, 092003 (2015).
\bibitem{Ablikim:2016qzw}
   M.~Ablikim {\it et al.} (BESIII Collaboration),
   Phys.\ Rev.\ Lett.\ {\bf 118}, 092001 (2017).
\bibitem{BESIII:2016adj}
   M.~Ablikim {\it et al.} (BESIII Collaboration),
   Phys.\ Rev.\ Lett.\ {\bf 118}, 092002 (2017).
%37


\bibitem{Ablikim:2018vxx}
   M.~Ablikim {\it et al.} (BESIII Collaboration)
   Phys.\ Rev.\ Lett.\ {\bf 122}, 102002 (2019).
\bibitem{Ablikim:2017oaf}
   M.~Ablikim {\it et al.} (BESIII Collaboration)
   Phys.\ Rev.\ Lett.\ {\bf 96}, 032004 (2017).
\bibitem{Wang:2014hta}
   X.~L.~Wang {\it et al.} (Belle Collaboration),
   Phys.\ Rev.\ D {\bf 91}, 112007 (2015).
\bibitem{Lees:2012pv}
   J.~P.~Lees {\it et al.} (BABAR Collaboration),
   Phys.\ Rev.\ D {\bf 89}, 111103 (2014).
\bibitem{Ablikim:2015uix}
   M.~Ablikim {\it et al.} (BESIII Collaboration),
   Phys.\ Rev.\ D {\bf 93}, 011102 (2016).

\bibitem{Ablikim:2019apl}
   M.~Ablikim {\it et al.} (BESIII Collaboration),
   Phys.\ Rev.\ D {\bf 99}, 091103 (2019).
\bibitem{Ablikim:2015xhk}
   M.~Ablikim {\it et al.} (BESIII Collaboration),
   Phys.\ Rev.\ D {\bf 91}, 112005 (2015).
\bibitem{Wang:2012bgc}
   X.~L.~Wang, X.~L.~Han, C.~Z.~Yuan, C.~P.~Shen, and P.~Wang,
   Phys.\ Rev.\ D {\bf 87}, 051101 (2013).
\bibitem{Ds*Ds*}
   W.~M.~Song, Ph.D thesis, {``$Study$ $of$ $the$ $production$ $and$ $decay$ $of$ $the$ $charmed$ $hadron$
  $at$ $BESIII$"}, Beijing IHEP, CAS, 2015.
  http://www.wanfangdata.com.cn/details/detail.do?{\_ }type=degree{\&}id=Y2957455

\bibitem{ZcGong} Q.~R.~Gong, $et$ $al.$, Phys.\ Rev.\ D {\bf 94} 114019 (2016).
\bibitem{ZcPang} Q.~R.~Gong, J.~L.~Pang, Y.~F.~Wang, H.~Q.~Zheng, Eur. Phys. J. C {\bf 78}, 276 (2018).
%\bibitem{Pakhlova:2008zza}
%   G.~Pakhlova {\it et al.} (Belle Collaboration),
%   Phys.\ Rev.\ D {\bf 77}, 011103 (2008).

\bibitem{Ablikim:2013wzq}
   M.~Ablikim {\it et al.} (BESIII Collaboration),
   Phys.\ Rev.\ Lett.\ {\bf 111}, 242001 (2013).

\bibitem{Voloshin:2012dk}
   M.~B.~Voloshin,
   Phys.\ Rev.\ D {\bf 85}, 034024 (2012).
%42


%\bibitem{Yuan:2005dr}
%   C.~Z.~Yuan, P.~Wang, and X.~H.~Mo,
%   Phys.\ Lett.\ B {\bf 634} (2006) 399-402.
%

\bibitem{Mo:2006ss}
  X.~H.~Mo, G.~Li, C.~Z.~Yuan, K.~L.~He, H.~M.~Hu, J.~H.~Hu, P.~Wang and Z.~Y.~Wang,
   Phys.\ Lett.\ B {\bf 640}, 182 (2006).
\bibitem{Bai:1999pk}
  J.~Z.~Bai {\it et al.} (BESII Collaboration),
  %``Measurement of the total cross-section for hadronic production by e+ e- annihilation at energies between 2.6-GeV - 5-GeV,''
 Phys.\ Rev.\ Lett.\ {\bf 84}, 594 (2000).
\bibitem{Bai:2001ct}
  J.~Z.~Bai {\it et al.} (BES Collaboration),
  %``Measurements of the cross-section for e+ e ---> hadrons at center-of-mass energies from 2-GeV to 5-GeV,''
 Phys.\ Rev.\ Lett.\  {\bf 88}, 101802 (2002)

\bibitem{morgan}D. Morgan, Nucl. Phys.  A{\bf 543},  632 (1992).
\bibitem{Cao:2019wwt}Q.~F.~Cao,  H.~R.~Qi, Y.~F.~Wang, H.~Q.~Zheng, Phys.\ Rev.\ D{\bf100},   054040 (2019).
%\bibitem{Gong:2016hlt}Q.~R.~Gong $et$ $al.$, Phys. Rev. D{\bf 94},  114019 (2016).
\bibitem{Meng:2014ota}C.~Meng $et$ $al.$, Phys.\ Rev.\ D{\bf 92},   034020 (2015).
\bibitem{Zhang:2009bv}O.~Zhang, C.~Meng, H.~Q.~Zheng, Phys.\ Lett.\ B{\bf 680}, 453 (2009).


\bibitem{BESIII:2020peo}
 M.~Ablikim {\it et al.} (BESIII Collaboration),
 arXiv:2007.12872.
\end{thebibliography}
\end{document}